\documentclass[twocolumn,showpacs,superscriptaddress,amsmath,amssymb,aps,prb]{revtex4}
\pdfoutput=1

\usepackage{graphicx}
\usepackage{dcolumn}
\usepackage{bm}

\begin{document}

\preprint{APS/123-QED}

\title{Tunable optical Aharonov-Bohm effect in a semiconductor quantum ring}

\author{Bin Li}
 \email{phymilky@gmail.com}
 \affiliation{Departement Fysica, Universiteit Antwerpen, Groenenborgerlaan 171, B-2020 Antwerpen, Belgium.}
  \author{F. M. Peeters}
 \email{Francois.Peeters@ua.ac.be}
 \affiliation{Departement Fysica, Universiteit Antwerpen, Groenenborgerlaan 171, B-2020 Antwerpen, Belgium.}

\date{\today}

\begin{abstract}
By applying an electric field perpendicular to a semiconductor quantum ring we show that it is possible to modify the single particle wave function between quantum dot (QD)-like to ring-like. The constraints on the geometrical parameters of the quantum ring to realize such a transition are derived. With such a perpendicular electric field we are able to tune the Aharanov-Bohm (AB) effect for both single particles and for excitons. The tunability is in both the strength of the AB-effect as well as in its periodicity. We also investigate the strain induce potential inside the self assembled  quantum ring and the effect of the strain on the AB effect.
\end{abstract}

\pacs{71.35.Ji, 73.21.La, 78.20.Bh, 78.20.Ls}
\maketitle

\section{\label{sec:introduction}Introduction}
 The electron and hole wave functions acquire an extra phase when moving in the presence of a perpendicular magnetic flux. The acquired extra phase will be different for electrons and holes, and this phase difference can be observed through photoluminescence (PL) experiments, which exhibits an optial Aharonov-Bohm (AB) effect. Experimental verification of this effect has been reported in PL measurements of radially polarized neutral excitons in a type-II quantum dot structure~\cite{ribeiro,tadic1}. However, the optical AB effects will be strongly suppressed~\cite{romer} when both the electron and the hole are spatially confined within the same region due to the Coulomb interaction, i.e. in a quantum ring (type-I quantum dot). This has spurred a considerable search and study of the optical AB effect of neutral excitons in semiconductor quantum rings by both theoretical~\cite{romer,jakyoung,huihu,govorov1,govorov2,silva,maslov}
 and experimental~\cite{ribeiro,kuroda} groups.

The optical AB effect in a quantum ring can be enhanced when the exciton is radially
polarized, either by the application of an external electric field~\cite{maslov,fischer,teodoro} or due to a radial asymmetry
in the effective confinement for electrons and holes~\cite{govorov1,govorov2,silva}. Theoretical studies~\cite{romer,govorov1,silva,chaplik}
on one dimensional rings predict that for a quantum ring whose radial size is comparable to the exciton Bohr
radius (i.e. in the weakly bound regime), the ground state energy could display nonvanishing AB oscillations. However, numerical calculations
on two dimension narrow rings~\cite{jakyoung,huihu} showed that: 1) there is no observable AB oscillation in the exciton ground state
energy, but 2) AB oscillation can be present in some low-lying excited energy levels.~\cite{huihu}.

Here we study a nano-ring structure, with and without strain, which has a varying height in the radial direction as is often found in self-assembled quantum rings~\cite{kuroda,offermans,ding1}. Applying a strong electric field in the perpendicular direction polarizes the exciton. We will show that the AB effect is strongly affected by the shape of the nano-ring structure and the external electric field which has a large effect on the Coulomb interaction between the electron and the hole, especially for quantum rings with a large height. This gives us the possibility to tune not only the shape of the electron and hole wave function, but even the magnetic field induced oscillations in the exciton energy through the application of an external electric field.

This paper is organized as follows: we present the physical model in Sec.~\ref{sec:2}. In the first part of Sec.~\ref{sec:3}, we use the finite element method to calculate the electron and the hole wave functions and energies in unstrained semiconductor quantum rings. Subsequently we calculate the total energy of the exciton in the presence of Coulomb interaction by diagonalizing the total Hamiltonian in the space spanned by the product of the single particle states. We show the tunability of the AB effect for both the single particles and the exciton energy by applying a perpendicular external electric field. In Sec.~\ref{sec:4}, we will give the equivalent results for a strained semiconductor ring, and we will show how the strain affects the results. Our conclusions are presented in Sec. V.

\section{\label{sec:2}Model}
\begin{figure}
\includegraphics[width=8cm]{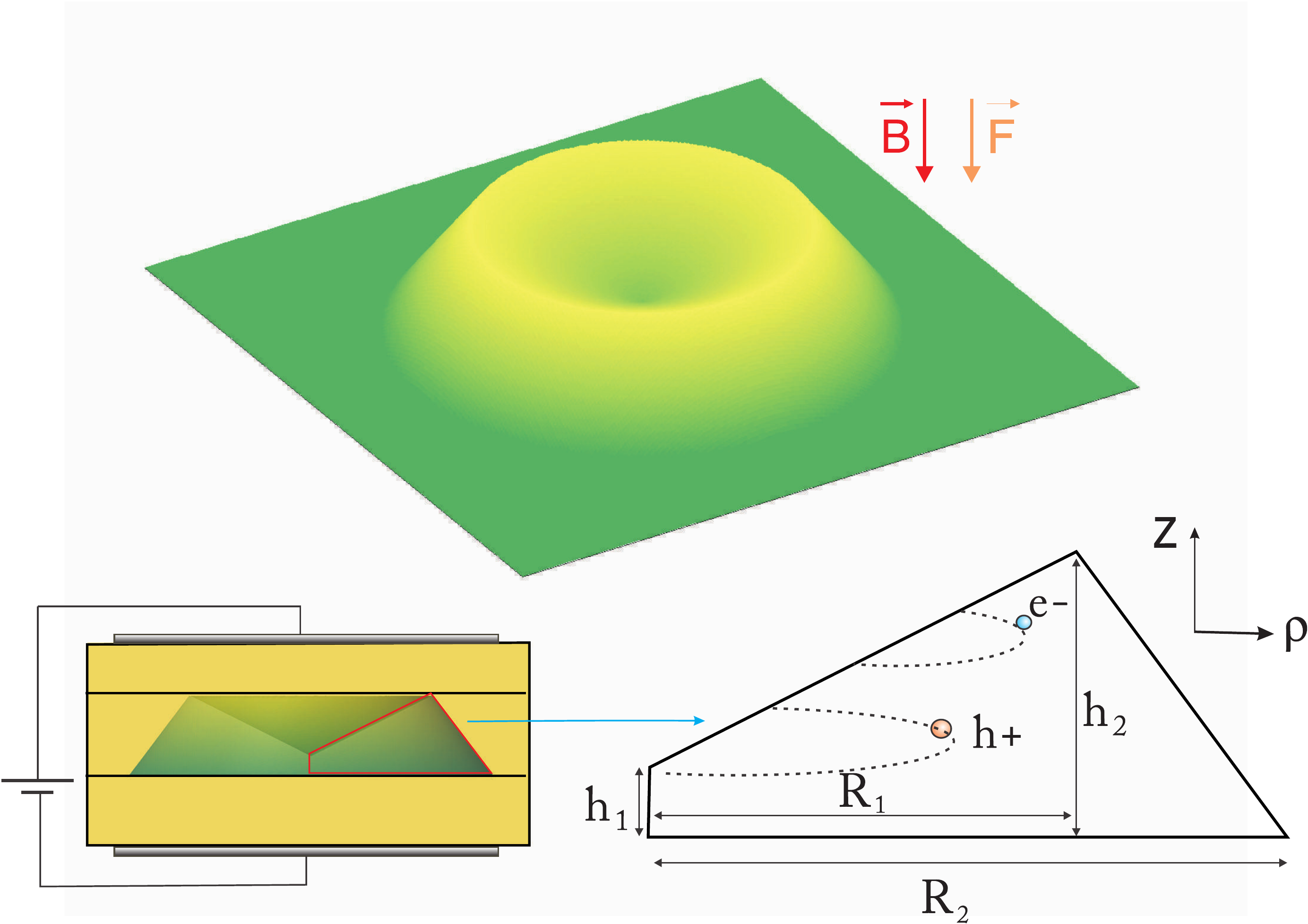}
\vspace{-0.2cm}\caption{\label{fig:model} (Color online) Schematics of the
volcano-like quantum ring. A perpendicular electric and magnetic field are applied in the vertical direction.  Bottom: the structure of the quantum ring in the $\rho$-$z$ plane.}
\end{figure}
Figure.~\ref{fig:model} shows the investigated geometry of the three dimensional ring, A volcano shaped ring is embedded in a barrier material which is different from the ring material. In a three dimensional semiconductor quantum ring, the full Hamiltonian of the exciton within the effective mass approximation is given by
\begin{eqnarray}
H_{tot}&=&\sum_{j=e,h}\left(\vec{P}_j-q_j\vec{A}_j\right)\frac{1}{2m_j}
\left(\vec{P}_j-q_j\vec{A}_j\right)+V_c\left(\vec{r}_e-\vec{r}_h\right)\nonumber\\
&&\!\! +\sum_{j=e,h}\delta
E_j(\vec{r}_j)+\sum_{j=e,h}V_j(\vec{r}_j)-eFz_e+eFz_h
 \label{eq:Hamil},
\end{eqnarray}
where $V_j(\vec{r}_j)$ is the confinement potential of the electron
(hole) due to the band offset of the two materials which will be different inside and outside the ring.
$V_c\left(\vec{r}_e-\vec{r}_h\right)=e^2/4\pi\varepsilon
|\left(\vec{r}_e-\vec{r}_h\right)|$ is the Coulomb potential between
the electron and the hole, and $\delta
E_j(\vec{r}_j)$ 
is the strain-induced shift of the energy which depends on the
strain tensor $\epsilon_{ij}$. We did not take the piezoelectric potential into account since it is
negligible compared to the other terms in our case. The last two
terms of Eq.~(\ref{eq:Hamil}) are the potential energy in the presence
of the perpendicular top to bottom directed electric field $F$.

\section{\label{sec:3}$GaAs/AlGaAs$ quantum ring}
\subsection{Single particle energy and wavefunction}
As a model system we consider first a volcano shaped GaAs ring surrounded by AlGaAs.  We assume that only the lowest electronic subband and the highest hole band (heavy hole) is occupied. In GaAs, the electron and the hole have effective mass $m_e/m_0=0.063$ and $m_h/m_0=0.51$, respectively. The static dielectric constant is $\varepsilon=12.5\varepsilon_0$ and the band gap is $E_g=1.42$ eV at helium temperature. While for AlGaAs, we have $m_e/m_0=0.082$, $m_h/m_0=0.568$, $\varepsilon=12.5\varepsilon_0$, and a band gap of $E_g=1.78$ eV. This results in a band gap difference of $\Delta E_g=360$ meV between GaAs and AlGaAs, which leads to a conduction band offset of about $\Delta E_c=250$ meV and a valence band offset of about $\Delta E_v=110$ meV, so we can take $V(\vec{r}_{e(h)})=0$ inside the ring and $\Delta E_{c(h)}$ outside the ring. The parameters used were taken from Ref.~\cite{handbook}. As the dielectric constant is practically the same everywhere and the difference of the lattice
constant for GaAs and AlGaAs is very small, we may ignore the dielectric mismatch effect, and the strain induced term in Eq.~(\ref{eq:Hamil}).

First, we calculate the single particle energy and the corresponding wave function. Because of cylindrical symmetry we rewrite the Hamiltonian in cylindrical coordinates. Assume the wave function of the single particles to be
$\Psi(\rho,z,\theta)=\psi_{e(h)}(\rho,z) e^{-i l_{e(h)} \theta}$,
after averaging out the angular part of the wave function, we obtain the 2D single particle Schr\"{o}dinger equation :
\begin{widetext}
\begin{equation}
\left(\!\!-\frac{\hbar^2}{2m_{e(h)}}\Big(\frac{\partial^2}{\partial
z^2}+\frac{\partial^2}{\partial
\rho^2}+\frac{1}{\rho}\frac{\partial}{\partial
\rho}-\frac{l_{e(h)}^2}{\rho^2}-\frac{q^2B^2\rho^2}{4m_{e(h)}}\Big)
\!-\!\frac{qBl_{e(h)}\hbar}{2m_{e(h)}}+V(\vec{r}_{e(h)})+qFz_{e(h)}\!\!\right)\!\psi_{e(h)}(\rho,z)
=E_{\rho,z}\psi_{e(h)}(\rho,z), \label{eq:schrodinger}
\end{equation}
\end{widetext}
here $q$ is $-e$ for the electron and $e$ for the hole. Taking advantage of cylindrical symmetry we only need to solve the problem in a half section of the ring as shown at the bottom inset of Fig.~\ref{fig:model}. We use the finite element methods to obtain the ground state energy for different angular moment $l_{e(h)}$ and electric field $F$ as a function of magnetic field $B$. For a better understanding and further consideration of the transition, we first write down the single particle Schr\"{o}dinger equation in dimensionless form (assume $B=0$ and $l_{e(h)=0}$):
\begin{widetext}
\begin{equation}
\left(-\frac{m_{e_0(h_0)}}{m_{e(h)}}\Big(\frac{\partial^2}{\partial
z^2}+\frac{\partial^2}{\partial
\rho^2}+\frac{1}{\rho}\frac{\partial}{\partial \rho}\Big)
+V^\prime(\vec{r}_{e(h)})+A_FF^\prime
z_{e(h)}\right)\psi_{e(h)}(\rho,z)=E^\prime_{\rho,z}\psi_{e(h)}(\rho,z),
\label{eq:schrodinger2}
\end{equation}
\end{widetext}
where $E^\prime_{\rho,z}=E_{\rho,z}/E_0$, and
$E_0=\hbar^2/2m_{e_0(h_0)}R_0^2$ is the energy unit. We take $R_0=1$ nm
as the length unit, and $m_{e_0(h_0)}$ is the effective mass of the
electron (hole) in GaAs. The effective band offset
$V^\prime(\vec{r}_{e(h)})=V(\vec{r}_{e(h)})/E_0$ is $0.425$ for
the electron and $1.472$ for the heavy hole. We take the unit of the
electric field to be $F_l=1$ kV$/$cm and the coefficient $A_F=F_l
R_0 e/E_0=0.00263m_{e_0(h_0)}$.

\begin{figure}
\includegraphics[width=8cm]{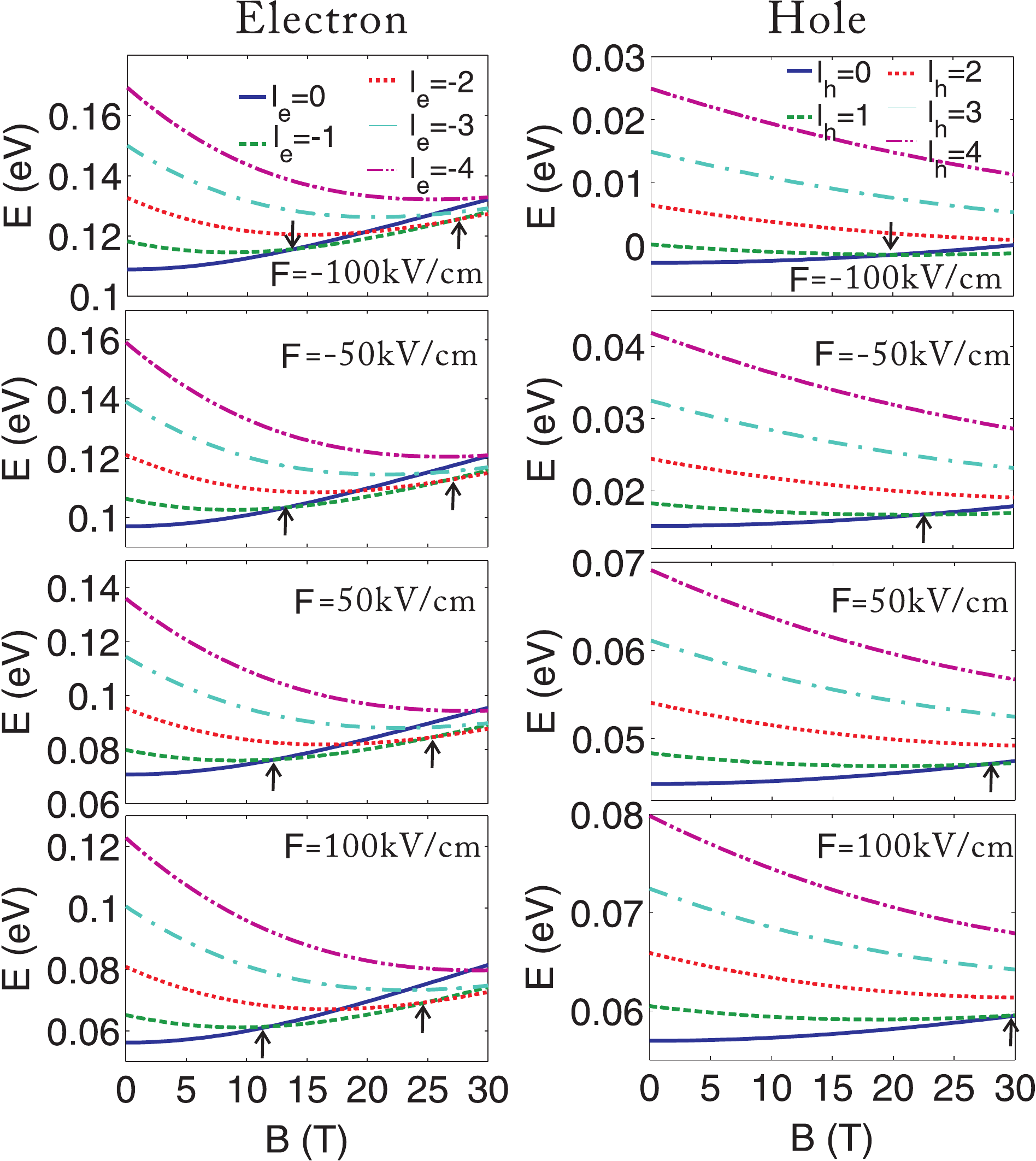}
\vspace{-0.2cm}\caption{\label{fig:volcaE}(Color online) Electron and hole energy for different electric field $F$ and angular moment $l_{e(h)}$ as a function of magnetic field. The size of the quantum ring is $h_1=4$ nm, $h_2=6$ nm, $R_1=12$ nm and $R_2=30$ nm. The arrows indicate angular momentum transitions in the single particle ground state.}
\end{figure}

Fig.~\ref{fig:volcaE} shows the single particle energies of the electron and the hole as a function of the perpendicular magnetic field $B$ for different values of the angular momentum and the electric field $F$. Here the size of the ring is chosen to be $h_1=4$ nm, $h_2=6$ nm, $R_1=12$ nm and $R_2=30$ nm. Notice that the absolute value of the angular momentum of the ground state for both the electron and the hole increases with $B$ and a pronounced AB oscillation is found (the second transition of the hole does not take place below $B=30T$, but it will definitely be there above 30T). It is clearly seen that the electric field has a larger effect on the hole than on the electron. Furthermore, the electric field pushes the electron and the hole wavefunction in opposite z-direction which results in a difference in the shape of the wave functions. When we decrease the electric field or even change the direction of the electric field, the angular momentum transition of the electron ground state will be more obvious, while for the hole, the transition becomes invisible (when $F=100$ kV/cm, there is no transition for the hole ground state for a magnetic field B below $30$ T). This will provide us with the possibility to tune the Aharonov-Bohm effect using the perpendicular electric field $F$, i.e. we can turn the AB effect off and on with this perpendicular electric field.
We can have a ring-like wave function (i.e. the probability is zero in the center of the ring) or a $QD$-like wave function (i.e. not zero probability of the wave function in the center) depending on the perpendicular electric field $F$.

\begin{figure}
\includegraphics[width=7cm]{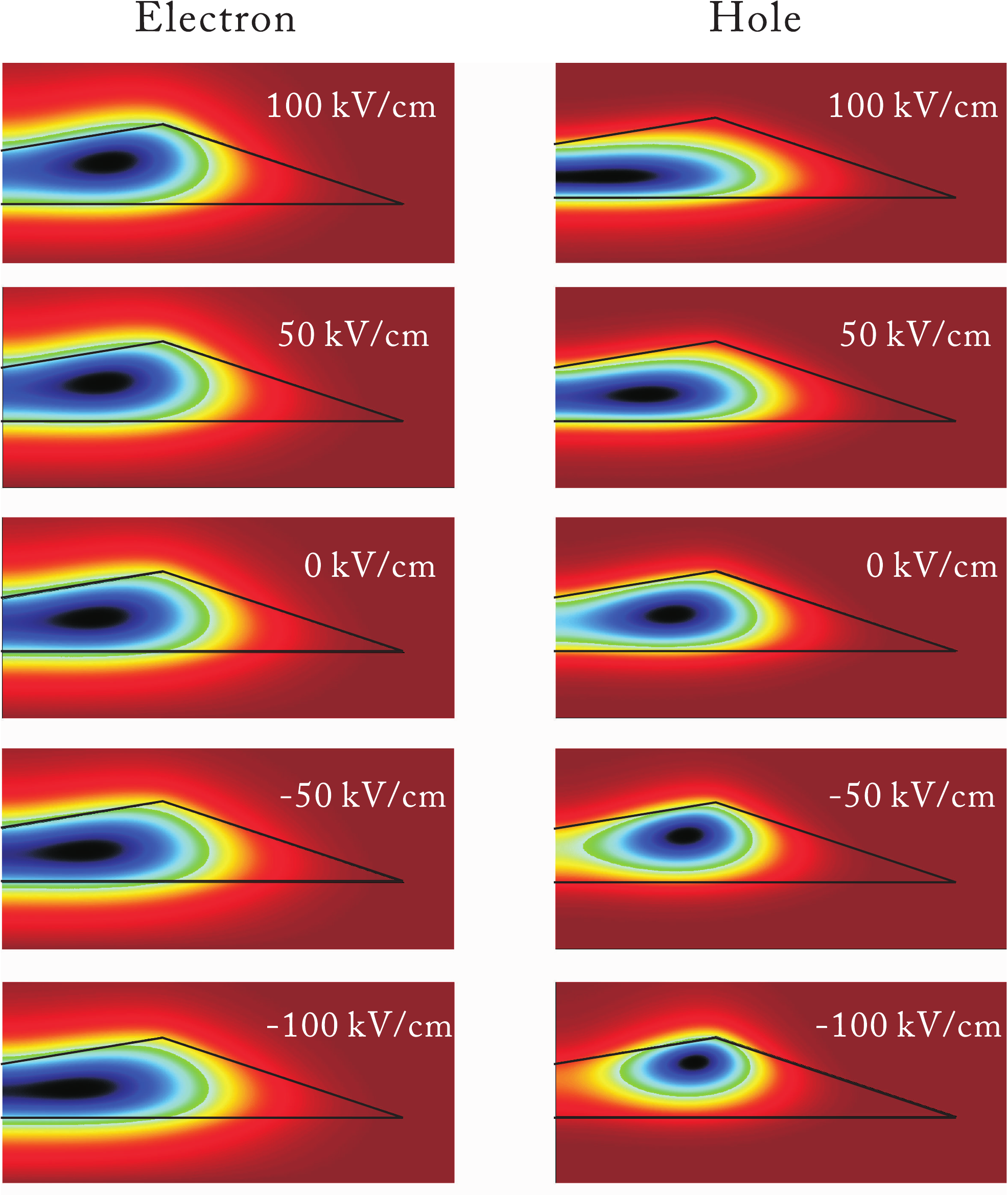}
\vspace{-0.2cm}\caption{\label{fig:contourwaveE2}(Color online) Contour plot of the ground state wave function of the electron and the hole for different values of the perpendicular electric field $F$ in the $(\rho,z)$ plane. Here $B=0$, $l_{e(h)}=0$ and the electric field $F$ from top to bottom are $100$ kV$/$cm, $50$ kV$/$cm, $0$, $-50$ kV$/$cm, $-100$ kV$/$cm. The contour of the ring is shown by the black curve.}
\end{figure}
Fig.~\ref{fig:contourwaveE2} shows the contour plot of the electron and the hole wave function in the $\rho-z$ plane. When the electric field changes from $100$ kV$/$cm to $-100$ kV$/$cm the wave function of the hole changes from a $QD$-like to a ring-like wave function, while the wave function of the electron changes from ring-like to $QD$-like. Notice that the wave function of the hole is more sensitive to the electric field, which can be understood from the dimensionless form of the Schr\"{o}dinger equation, Eq.~(\ref{eq:schrodinger2}). Notice that the term related to the electric field is $0.00263m_{e_0(h_0)}F^\prime z$, which is proportional to the effective mass. As the heavy hole has a larger effective mass, the effect of the electric field will be larger.

\begin{figure}
\includegraphics[width=7.5cm]{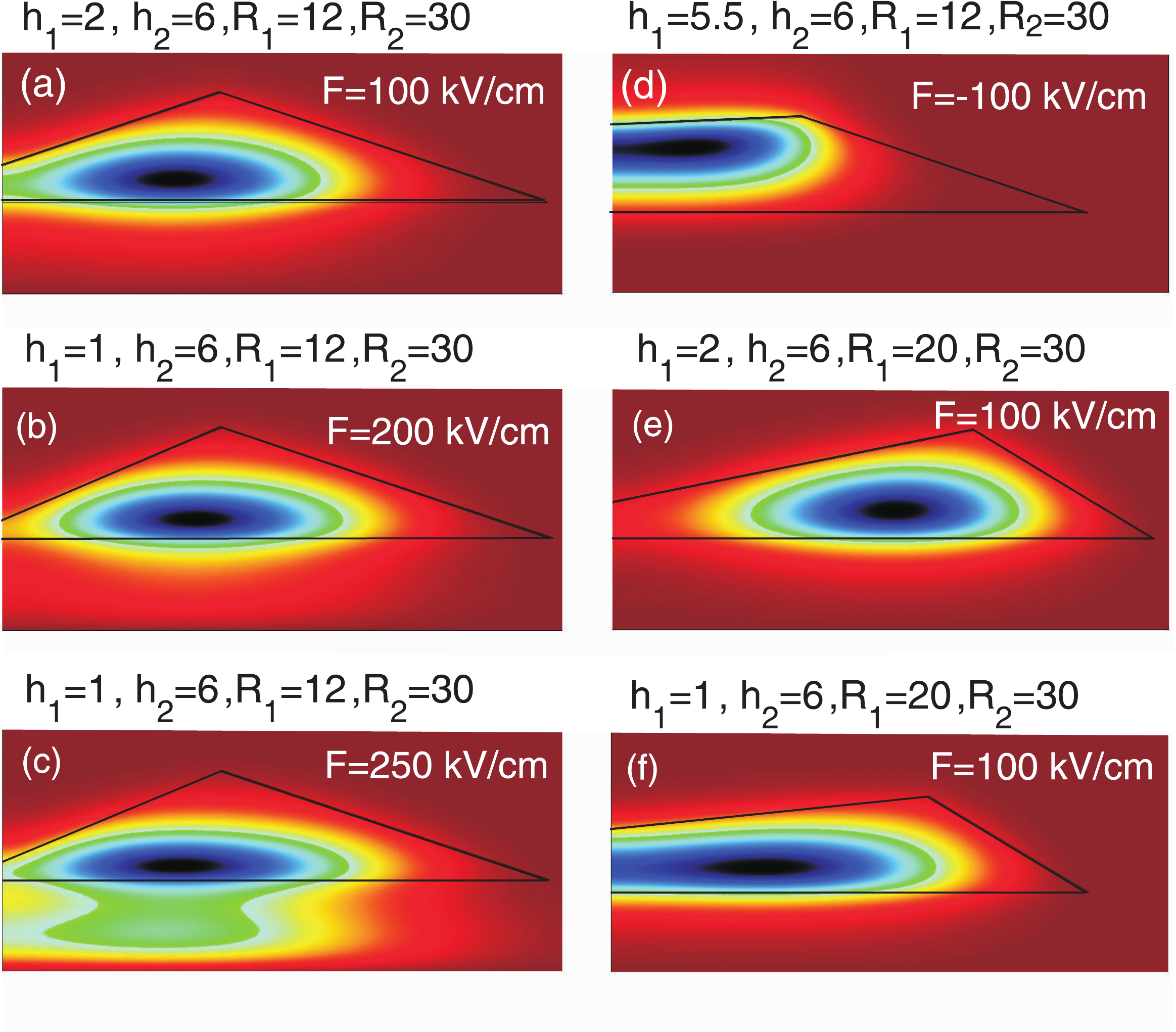}
\vspace{-0.2cm}\caption{\label{fig:contourwaveE3}(Color online) Contour plot of
the ground state wave function of the heavy hole for different sizes of
the ring. The size (in unit of nm) and the corresponding electric field are specified
above each figure.}
\end{figure}

From a careful inspection of Eq.~(\ref{eq:schrodinger2}) and Fig.~\ref{fig:contourwaveE2} we determine the constraints in order to have an electric field tunable Aharonov-Bohm effect: 1) The electric field term $A_FF^\prime dz$ must be comparable to the difference of the confinement energy of the $QD$-like and ring-like wave function, which is proportional to
$1/h_1^2-1/h_2^2$. When passing from the $QD$-like to the ring-like wave function, the difference of the $z$ coordinate is proportional to $(h_2-h_1)/2$, so the first constraint condition is $A_FF^\prime(h_2-h_1)/2\sim1/h_1^2-1/h_2^2$. For example in our case, $h_1=4$ nm, $h_2=6$ nm and for the heavy hole with an effective mass $m_h=0.51m_0$ the electric field should be at least $2(1/h_1^2-1/h_2^2)/A_F/(h_2-h_1)*F_l = 2(h_1+h_2)/(h_1^2h_2^2A_F)*F_l$, which is several tens of kV$/$cm. If we decrease $h_1$, then the electric field should be larger. We see from Fig.~\ref{fig:contourwaveE3}(a) that if $h_1=2$ nm, the wave function of the hole is still ring-like when $F=100$ kV$/$cm, and we need to increase the electric field in order to have a $QD$-like wave function, which leads to the second constraint condition: 2) the confinement energy of the electron (hole) when it is $QD$-like (i.e. which is $\sim1/h_1^2$) should be smaller than the band offset $V^\prime(\vec{r}_{e(h)})$ of the quantum ring, otherwise the electron (hole) will move out of the ring to the barrier for large electric field values, as can been seen from Figs.~\ref{fig:contourwaveE3}(b,c). The wave function is still ring like when $F=200$ kV$/$cm, but when we increase the electric field to $F=250$ kV$/$cm, the wave function starts to penetrate in the barrier region. The AB effect is still tunable, but the effect of the electric field will be very small in case of a small ring height $h_2$. 3) The confinement energy difference of the $QD$-like and $ring-like$ states, which are proportional to $1/h_1^2-1/h_2^2$ should not be very small, otherwise the wave function of the electron (hole) will always be $QD$-like and shows no observable AB effect (as we can see from Fig.~\ref{fig:contourwaveE3}(d), when $h_1=5.5$ nm and $h_2=6$ nm the wave function is still $QD$-like when $F=-100$ kV$/$cm). 4) In order to have a $QD$-like wave function, the radius of the ring $R_1$ should not be very large compared to the height of the ring (otherwise the electric field will not have a significant effect on the wave function), and this constraint is much more important when the height of the ring is small (we can see this by comparing Figs.~\ref{fig:contourwaveE3}(e, f) with Figs.~\ref{fig:contourwaveE3}(a) and ~\ref{fig:contourwaveE2}). The above four constraints were obtained for holes, but they are also valid for electrons.

\subsection{Exciton energy including Coulomb interaction}
The total exciton energy is calculated using the
configuration interaction (CI) method. We first construct the total
exciton wave function as a product of linear combinations of single electron
and hole wave functions, and then calculate the matrix elements of
the exciton Hamiltonian. By diagonalizing the obtained matrix we obtain the exciton energies and their corresponding wave functions.

The Hamiltonian of the exciton is
\begin{equation}
H_{tot}=H_e+H_h+U_c,
\end{equation}
here $H_e$ and $H_h$ are the single particle Hamiltonians,
\begin{equation}
U_c=U_0*\frac{1}{\sqrt{{\rho^\prime_e}^2+{\rho^\prime_h}^2-2\rho^\prime_e\rho^\prime_h\cos(\varphi_e-\varphi_h)+\left(z^\prime_e-z^\prime_h\right)^2}},
\end{equation}
is the Coulomb energy between the electron and the hole, where $U_0=-e^2/4\pi\varepsilon R_0$.

As the total angular momentum is a good quantum number because of cylindrical symmetry, we assume the wave function of the exciton to be:
\begin{equation}
\Psi_L\left(\vec{r}_e,\vec{r}_h\right)=\sum_{k}C_k
\Phi_k\left(\vec{r}_e,\vec{r}_h\right),
\label{eq:totalwavef}
\end{equation}
for given total angular moment $L$. Here, $\Phi_k\left(\vec{r}_e,\vec{r}_h\right)=\psi_{n_e}e^{-i l_e\varphi_e}\psi_{n_h} e^{-il_h \varphi_h}$, and the exciton wavefunction is constructed out of single particle eigenstates, and $k$ stands for the indices $(n_e,n_h,l_e,l_h)$. $l_e+l_h=L$ should always be satisfied for fixed $L$, $n_e (n_h)$ is the quantum number of the single particle radial wave function. As the energy of the single particle eigenstates with large angular moment quantum number $l_e$ ($l_h$) and quantum number $n_e$ ($n_h$) is much larger as compared to the ones with smaller angular moment quantum number, we can limit ourselves to several tens of single particle eigenstates. With this wave function, we can construct the matrix of the total Hamiltonian and by diagonalizing the obtained matrix, we can get the eigenvalues and eigenvectors.

\begin{figure}
\includegraphics[width=6.6cm]{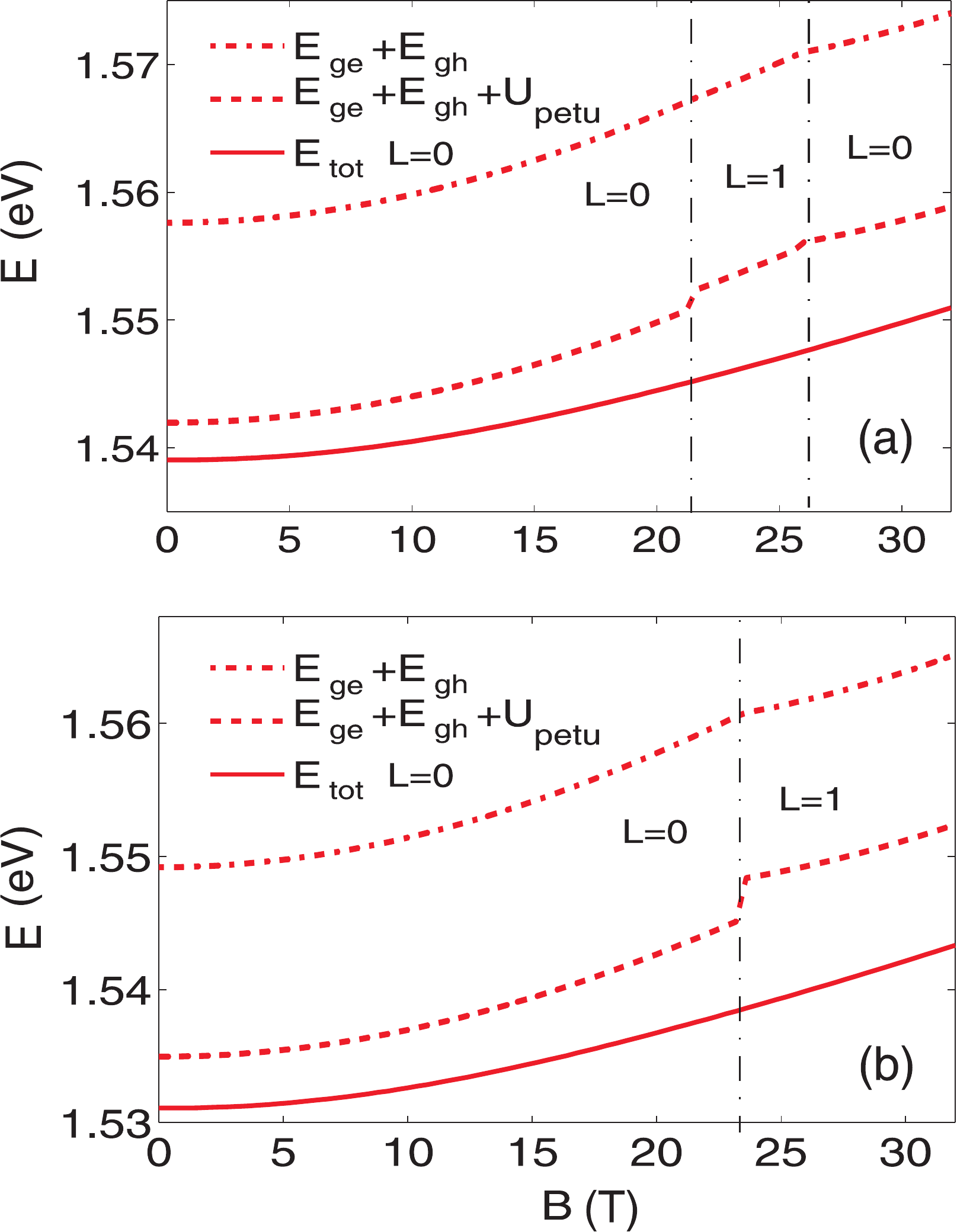}
\vspace{-0.2cm}\caption{\label{fig:etotc} Exciton energy for total angular momentum $L=0$ as obtained from the CI method (solid line) and compared to the results of the ground exciton energy without Coulomb interaction (dot-dashed line), and the ground state energy by taking the Coulomb energy as a perturbation (dashed line). Here the top figure is for $F=0$, and the bottom one is for $F=150$ kV/cm.}
\end{figure}

For a fixed total angular moment, we have
\begin{widetext}
\begin{equation}
U_c(k,j)=C_kC_j\left<\Phi_k\left(\vec{r}_e,\vec{r}_h\right)|U_c|\Phi_j\left(\vec{r}_e,\vec{r}_h\right)\right>
      =C_kC_j
      \int\int\int\int\psi_{n_e}\psi_{n_h}
      \psi_{m_e}\psi_{m_h} A_{{l_e}_j,{l_h}_j,{l_e}_k,{l_h}_k} \rho^\prime_e
      d\rho^\prime_e \rho^\prime_h d\rho^\prime_h dz^\prime_e
      dz^\prime_h ,
\end{equation}
and where $\textrm{Em}$ is the angular part of the integral
\begin{equation}
A_{{l_e}_j,{l_h}_j,{l_e}_k,{l_h}_k}=U_0\int_0^{2\pi}\int_0^{2\pi}d\varphi_e
d\varphi_h\frac{e^{-i({l_e}_j+{l_h}_j-{l_e}_k-{l_h}_k)\varphi_e}                                                                                                                       e^{i({l_h}_j-{l_h}_k)(\varphi_e-\varphi_h)}}{\sqrt{{\rho^\prime_e}^2
+{\rho^\prime_h}^2-2\rho^\prime_e\rho^\prime_h\cos(\varphi_e-\varphi_h)+\left(z^\prime_e-z^\prime_h\right)^2}} .
\label{eq:Em}
\end{equation}
\end{widetext}
We know from Eq.~(\ref{eq:Em}) that, in order to make the integral nonzero, ${l_e}_j+{l_h}_j-{l_e}_k-{l_h}_k$ should be zero, which indicates that the total angular moment is a good quantum number. If we take $\varphi_e-\varphi_h$ as a new variable in Eq.~(\ref{eq:Em}), it can be changed into an elliptic integral, which is easy to calculate numerically. The remaining integral is easy to calculate numerically. The effective exciton Bohr radius is $a_B=4\pi\varepsilon \hbar^2/(\mu e^2)=11.7673$ nm ($\mu=1/(1/m_e+1/m_h)$), and we take a small volcano shaped ring ($r_c=10$ nm, $R=16$ nm, $h_1=2$ nm and $h_2=4$ nm), whose size is comparable to the exciton Bohr radius. The results of the total exciton energy for $F=0$ and $F=150$ kV/cm are shown in Fig.~\ref{fig:etotc}, and as a comparison, we also show the results of the ground exciton energy without Coulomb interaction and the total ground state energy by taking the Coulomb energy as a perturbation (we took only the lowest electron and hole state into account, the result is much closer to the real one for a smaller ring when the Coulomb energy is small as compared to the kinetic energy). From Fig.~\ref{fig:etotc} we know that the Coulomb interaction has a large effect on the total exciton energy, and unlike the perturbation result, the exciton energy from the CI method shows a monotonous increase with magnetic field. There are no oscillations as function of the magnetic field. The existence of the top to bottom directed electric field decreases the exciton energy by tens of meV but does not bring any oscillation.

\begin{figure}
\includegraphics[width=7.5cm]{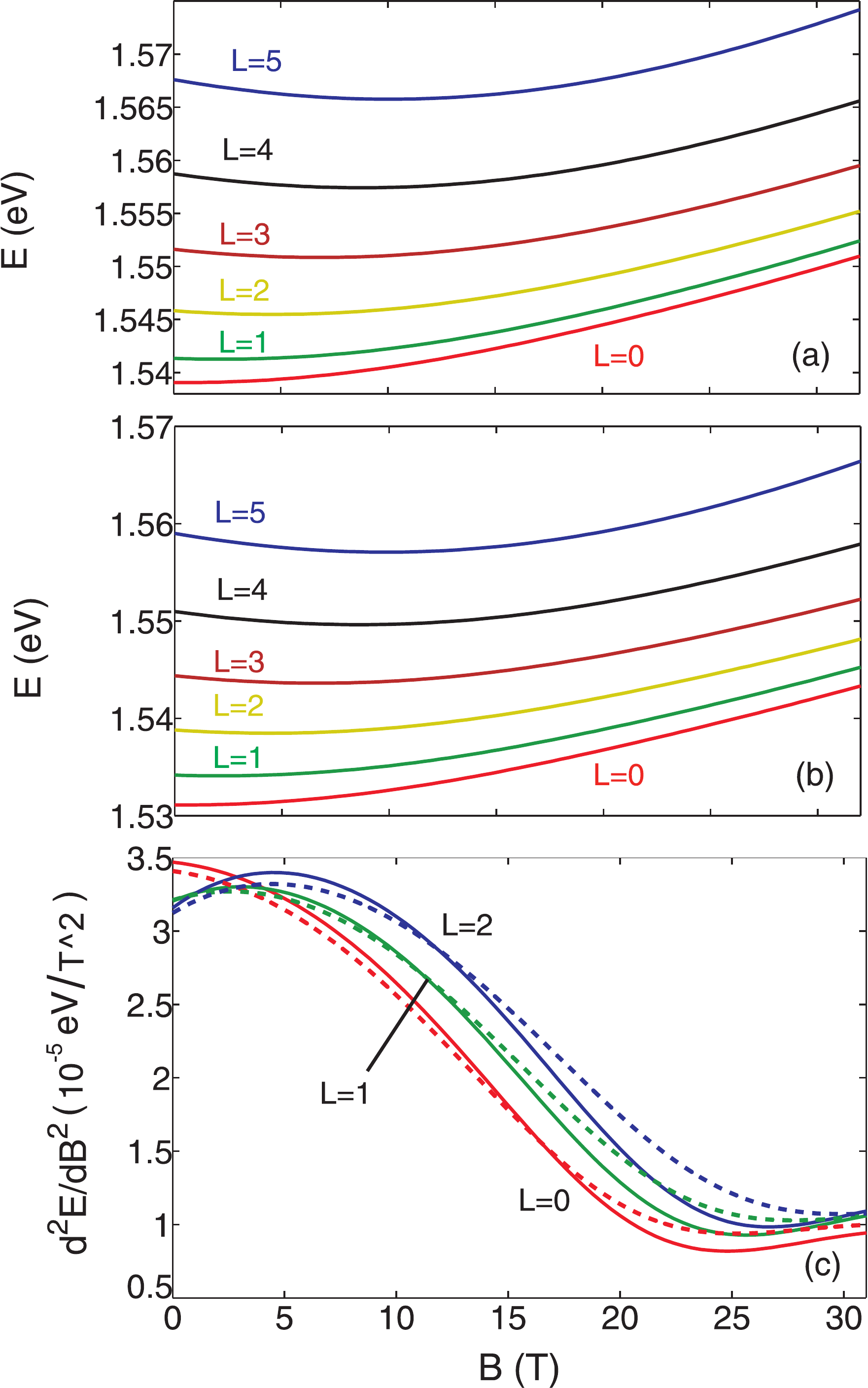}
\vspace{-0.2cm}\caption{\label{fig:excitot}(Color online) Ground state energy and the five lowest excited exciton energies as a function
of the magnetic field. The top figure is for the case of $F=0$, and the middle one is for $F=150$ kV/cm. (c) Second
derivative of the exciton energy with respect to the magnetic field.}
\end{figure}

Figures \ref{fig:excitot}(a, b) show the CI results, for the exciton energies with different values of the total angular momentum $L$, when $F=0$ and $F=150$ keV. It shows that the state with total angular moment $L=0$ is always the ground state, which is different from the perturbation theory result shown in Fig.~\ref{fig:etotc} where the total angular moment of the ground state switches between $L=0$ and $L=1$. We notice that the energy of the exciton states show no oscillation with increasing magnetic field, both with or without perpendicular electric field. The effect of the electric field on the exciton leads to a decrease of the exciton energy and a larger difference between the energy of the lowest exciton states. In Fig.~\ref{fig:excitot}(c) we plot the second derivative of the exciton energy with respect to the magnetic field, for both with (solid curves) and without (dotted curves) electric field. We find that the second derivative of the exciton energy shows a clear oscillation with respect to the magnetic field, both for the ground state and excited states energy. But since the size of the ring, as we specified above, is very small, the period of the oscillation is large. The oscillation in $d^2E_{tot}/dB^2$ is slightly more pronounced when a perpendicular electric field is applied. The reason is that the electric field makes the electron and the hole move in the opposite directions which enhances the polarity of the exciton. 


\begin{figure}
\includegraphics[width=8.5cm]{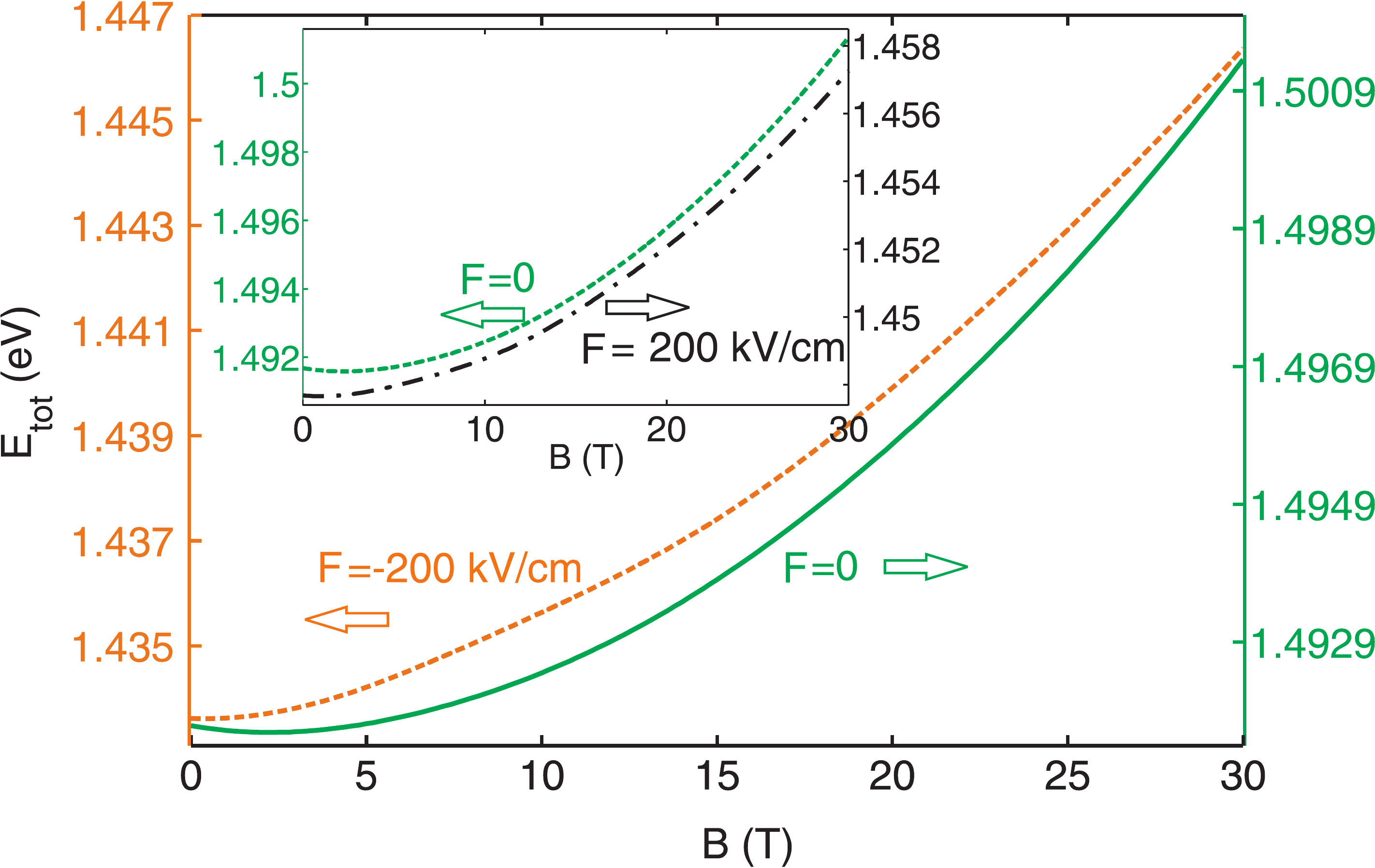}
\vspace{-0.2cm}\caption{\label{fig:Etotl0}(Color online) Exciton ground state energy as a function of the magnetic field for $F=0$, $F=-200$ kV/cm and $F=200$ kV/cm (inset). The curves in different color stand for different $y$ scale.}
\end{figure}
Till now, we did not observe any optical AB effect in the exciton ground state energy and any large influence of the perpendicular electric field on the oscillation. In order to see the effect of the electric field more clearly, we calculate the energy for a quantum ring with a larger height and a much larger electric field. Since the period of the oscillation for the previous case is large, we increase the radius of the ring in order to decrease the period. Fig.~\ref{fig:Etotl0}(a) shows the exciton ground state energy with respect to the magnetic field in the case of $F=0$ kV/cm, $F=-200$ kV/cm and $F=200$ kV/cm (inset of Fig.~\ref{fig:Etotl0}). Here $R_1=14$ nm, $R_2=18$ nm, $h_1=1$ nm and $h_2=10$ nm (the size is still comparable to the effective Bohr radius). As the state with total angular moment $L=0$ always has the lowest energy, the ground state is the same as the $L=0$ state here . It is clear that although the $F=0$ state shows no oscillation at all, the ground state energy in the presence of an electric field shows a weak but observable oscillation, especially for the case of $F=-200$ kV/cm. And the ground state energy is smaller in the presence of a top to bottom directed electric field. 

\begin{figure}
\includegraphics[width=8.5cm]{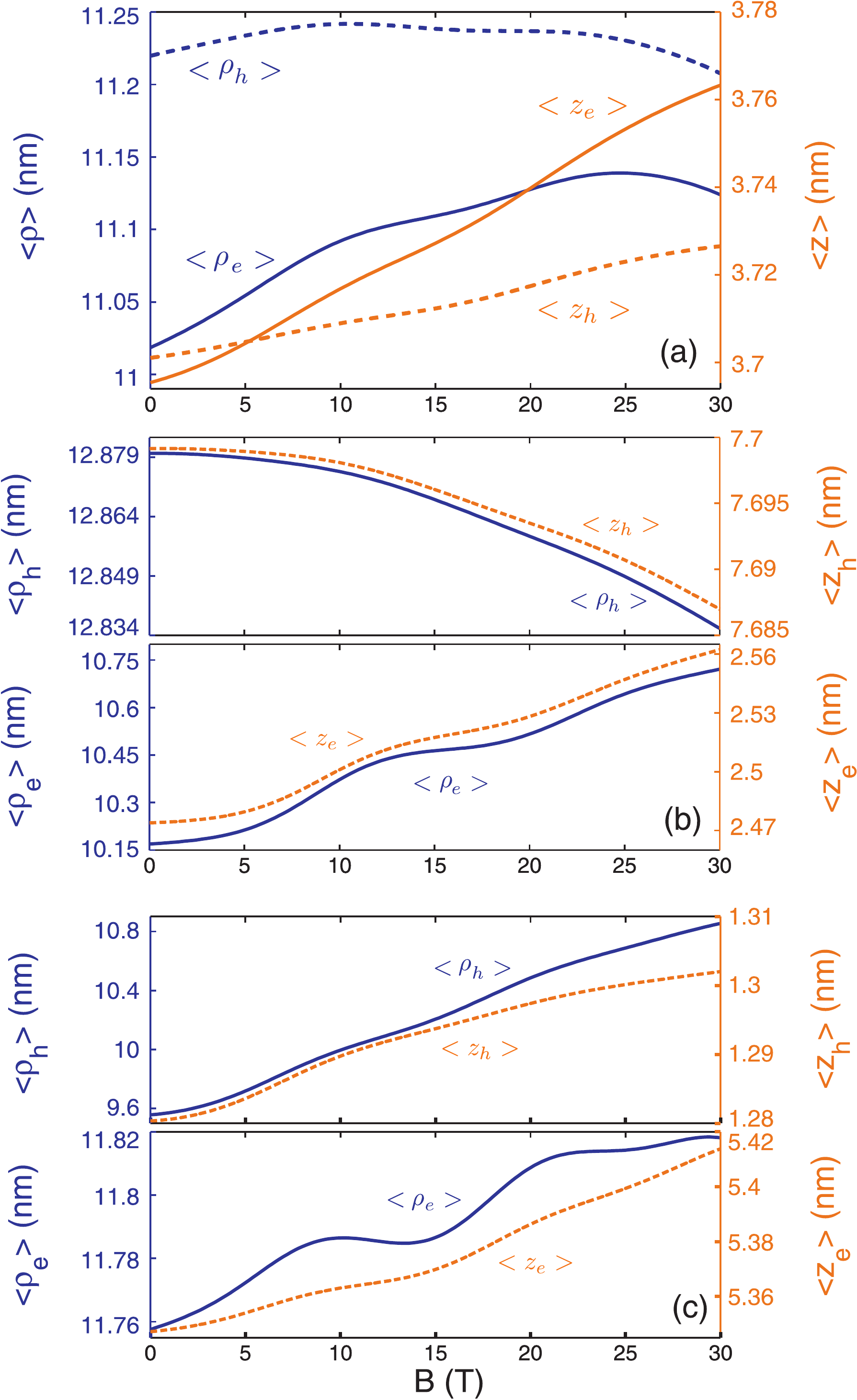}
\vspace{-0.2cm}\caption{\label{fig:averagezr}(Color online) Average distance $<\rho>$ (left $y$ scale in same color) and $<z>$ (right $y$ scale) in unit of nm of the electron (hole) as a function of magnetic field for (a) $F=0$, (b) $F=-200$ kV/cm and (c) $F=200$ kV/cm. In (b) and (c) the curves of the electron and the hole are separated in different sub figures.}
\end{figure}
For a better understanding of the existence of the ground state AB oscillations and the difference of the oscillation period for different values of the electric field, we calculate the average value $<\rho>$ in the radial direction and $<z>$ in the $z$ direction of the electron and the hole in the exciton ground state, which we show in Fig.~\ref{fig:averagezr}. From Fig.~\ref{fig:averagezr}(a) we see that $<\rho>$ and $<z>$ for both electron and hole have no clear step-like behavior, but they rather change smoothly. Furthermore, the difference of the average values of the electron and the hole in both radial and $z$ direction is rather small, especially in the $z$ direction. This makes then the Coulomb interaction energy very large and the polarity of the exciton extremely small, which greatly weakens the oscillations in the exciton ground state energy. That is the reason why we can not observe any oscillation of the exciton energy in the absence of the electric field. But in the presence of a perpendicular electric field, we see from Figs.~\ref{fig:averagezr}(b) and ~\ref{fig:averagezr}(c) that the difference of the average values between the electron and the hole becomes much larger. And the average values $<\rho>$ and $<z>$ show pronounced increasing (or decreasing) step-like behavior with increasing magnetic field, especially for the electron. In both cases of $F=200$ kV/cm and $F=-200$ kV/cm, $<\rho_e>$ and $<z_e>$ show obviously oscillations with increasing magnetic field, and their oscillation period is more or less the same as the period of the oscillation in the exciton energy. That is because the coupling of the electron and hole states with different angular momentum becomes much smaller as a result of the weak coulomb interaction. If we look at the normalization constant $C_k$ in Eq.~(\ref{eq:totalwavef}) we know that the step-like behavior in $<\rho_e>$ and $<z_e>$ in Figs.~\ref{fig:averagezr}(b,c) originate from the angular momentum transition of the main contributing basis function in the total exciton wave function. With increasing magnetic field, the angular momentum pair $(l_e,l_h)$ in the state $\Phi_k\left(\vec{r}_e,\vec{r}_h\right)$ which has the largest contribution to the total wave function changes from $(0,0)$ to $(-1,1)$ or to even larger values, and this transition becomes more notable in the presence of a perpendicular electric field as the coupling between the different contributing basis functions is much weaker. 

From Fig.~\ref{fig:averagezr} we also see that there are some differences between the cases $F=-200$ kV/cm and $F=200$ kV/cm. In the case of $F=-200$ kV/cm, the electron is in the center area of the ring while the wave function of the hole has a ring shape. The Coulomb interaction makes the electron (hole) move to the top (center) area of the ring towards the hole (electron). When the magnetic field is increased, the electron (hole) wave functions change towards the area with larger (smaller) values of $\rho$ in which the hole (electron) is located, but as a result of the strong electric field and confinement energy, only $<\rho_e>$ is mainly changed. Moreover, compared to the electron, the averages $<\rho>$ and $<z>$ for the hole change more smoothly with increasing magnetic field. The behavior for $F=200$ kV/cm is different. The electron now is in the top area of the ring, $\rho_e$ and $z_e$ change over a rather small range as a result of the strong confinement in the top area of the ring. Moreover, with increasing magnetic field,  $<\rho>$ and $<z>$ for both the electron and the hole increase towards the area with larger value of $\rho$, but the change for the electron is very small. This is different from the case of $F=-200$ kV/cm.
\begin{figure}
\includegraphics[width=8.5cm]{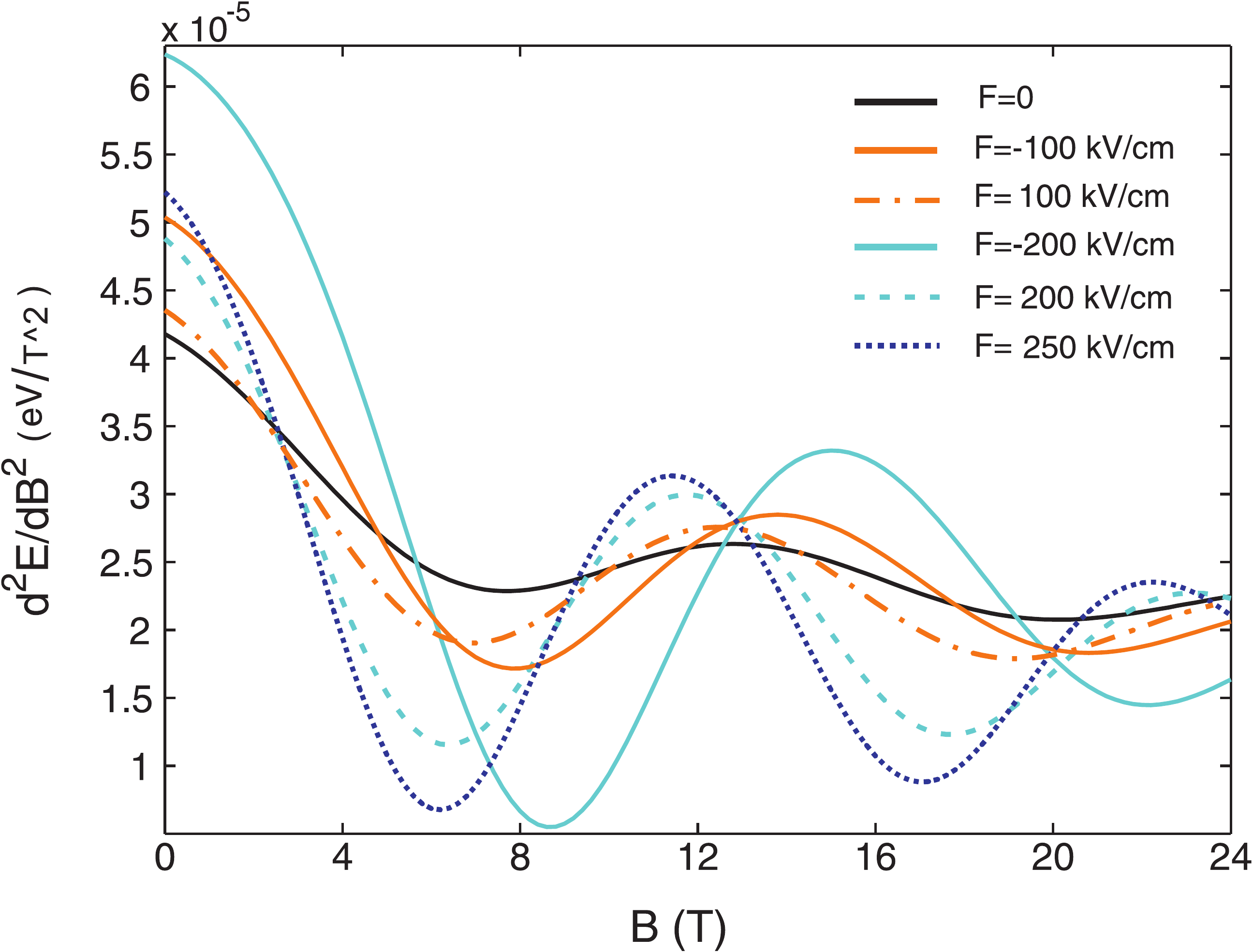}
\vspace{-0.2cm}\caption{\label{fig:2ndEl0}(Color online) Second derivative of the exciton ground state energy with respect to the magnetic field for different values of the perpendicular field $F=0$ (black solid line), $F=-100$ kV/cm (red solid line), $F=100$ kV/cm (red dash-dotted line), $F=-200$ kV/cm (light blue solid line), $F=200$ kV/cm (light blue dashed line) and $F=250$ kV/cm (dark blue dotted line)}
\end{figure}

In Fig.~\ref{fig:2ndEl0} we show the second derivative of the exciton ground state energy with respect to the magnetic field for different values of the perpendicular field. We clearly notice that the oscillation is enhanced in the presence of the perpendicular electric field, independent of the direction of the electric field. The oscillation is much larger when the electric field is directed from bottom to top. This can be understood by comparing Figs.~\ref{fig:averagezr}(b,c), where the $<\rho>$ ($<z>$) differences between the electron and the hole are both larger. But the difference is smaller in the case of $F=200$ kV/cm, where the Coulomb energy $U_c$ is larger, which weakens the AB oscillation in the exciton ground state energy as compared to the case of $F=-200$ kV/cm. We also verified in Fig.~\ref{fig:2ndEl0} that the period of the oscillation is larger when increasing the bottom to top directed electric field, but smaller by increasing the top to bottom directed electric field. The period change is a consequence of the change of average radius of the electron and/or hole with changing electric field. However, it is hard to obtain this result only from the changing in $<\rho_e>$ and $<\rho_h>$ alone. As we notice from Figs.~\ref{fig:averagezr}(b) and ~\ref{fig:averagezr}(c) that $<\rho_e>$ and $<z_e>$ is smaller in the case of $F=-200$ kV/cm as compared to $F=0$ kV/cm, but $<\rho_h>$ and $<z_h>$ is larger, which probably decreases the AB oscillation period of the total exciton energy. Moreover, although $<\rho_e>$ and $<z_e>$ is larger in the case of $F=200$ kV/cm, but $<\rho_h>$ and $<z_h>$ is smaller. The point is that as the hole has a much larger effective mass, the average radius of the exciton is mainly determined by the electron. As we explained in Appendix. B, the period of the oscillation is proportional to $m_h/<\rho_e>^2+m_e/<\rho_h>^2$, and is mainly determined by the effective radius of the electron.

In order to study the exciton recombination, we calculate the oscillator strength (the dimensionless quantity that expresses the strength of the transition) for the state with total angular momentum $L=0$. The oscillator strength for the exciton ground state is defined as\cite{garnett,tadic}
\begin{equation}
f_g=\frac{2}{m}\frac{|<ex|\sum_i p_{x{i}}|0>|^2}{E_{ex}-E_0},
\end{equation}
where $m$ is the free electron mass and $|ex>$ ($|0>$) is the exciton state (single electron and hole pair). By using the envelope-function approximation, we can derive\cite{henry}
\begin{equation}
f_g=\frac{2P^2}{m(E_{ex}-E_0)}\left|\int\Psi_g\left(\vec{r}_e,\vec{r}_e\right)d\vec{r}_e\right|^2,
\end{equation}
here $P$ includes all intra matrix-element effects, $\Psi_g\left(\vec{r}_h,\vec{r}_e\right)$ is the exciton ground state wave function. For simplicity, here we will focus on the main variable part of the oscillator strength which is $O_f=\left|\int\Psi_g\left(\vec{r}_e,\vec{r}_e\right)d\vec{r}_e\right|^2$, named the overlap integral.
\begin{figure}
\includegraphics[width=8cm]{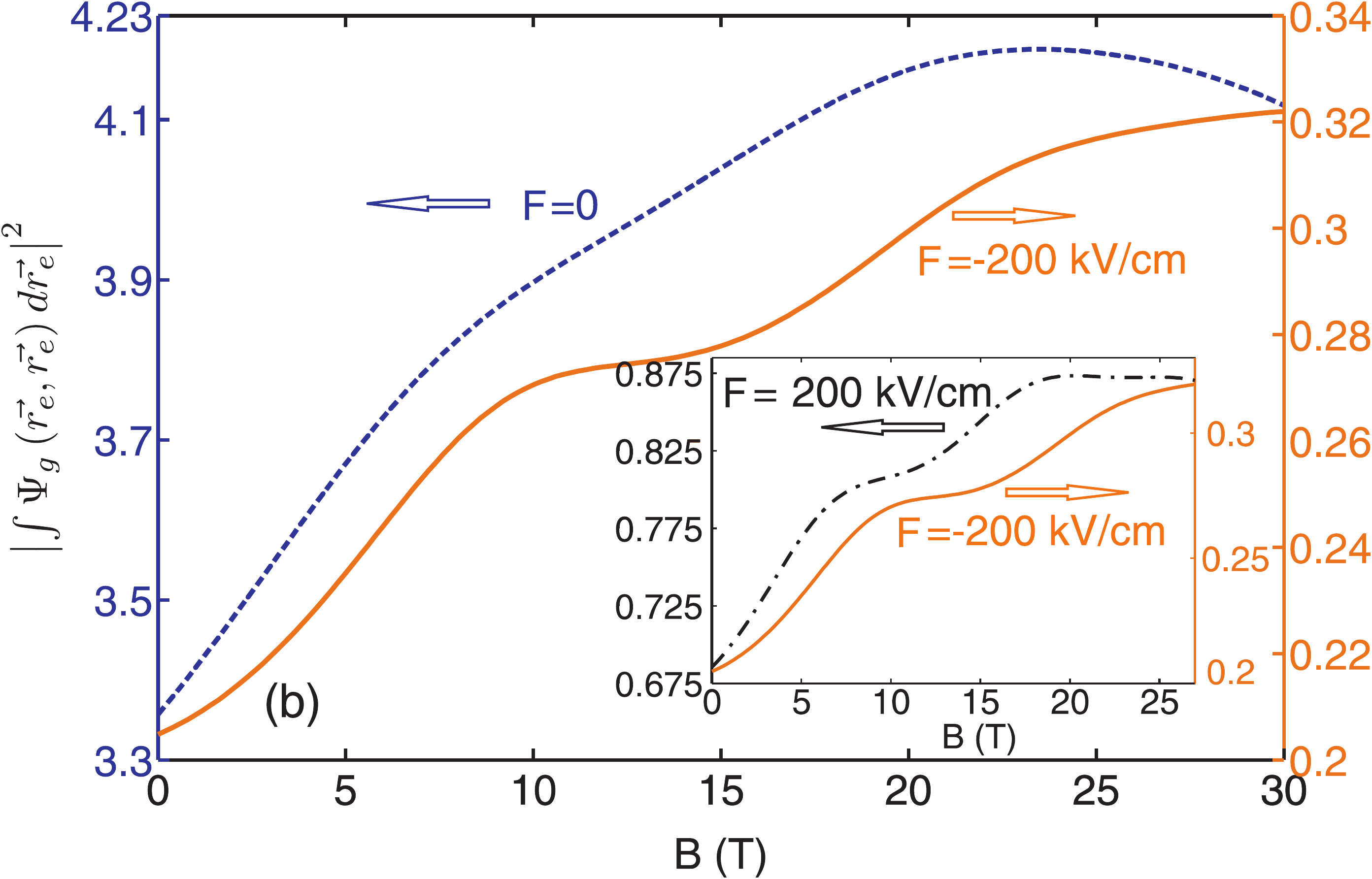}
\vspace{-0.2cm}\caption{\label{fig:oscill}(Color online) Overlap integral $O_f=\left|\int\Psi_g\left(\vec{r}_e,\vec{r}_e\right)d\vec{r}_e\right|^2$ of the exciton ground state energy as a function of the perpendicular magnetic field. The pink dashed line is for $F=0$, red solid line stand for $F=-200$ kV/cm and black dash-dotted line is for $F=200$ kV/cm. Note the different scales for the three curves}
\end{figure}
The result of the overlap integral for the exciton ground state is shown in Fig.~\ref{fig:oscill}, for $F=-200$ kV/cm, $F=0$ kV/cm and $F=200$ kV/cm. In the presence of the electric field, $O_f$ exhibits more structure, and we can clearly see steps in the overlap integral. The steps in the overlap integral have the same period as the AB oscillation, which also originates from the angular momentum transition. Moreover, the overlap integral for the case of $F=0$ is much larger, which is a consequence of the extremely small polarity of the exciton. $O_f$ in case of $F=-200$ kV/cm shows the smallest value as compared to the other two cases, which means the exciton has the largest polarity and is most stable, and the AB effect is the strongest (which confirm our results of Fig.~\ref{fig:2ndEl0}). We also find that for all the three cases, the overlap integral increases with increasing magnetic field, which means the exciton has a smaller polarity and also a weaker AB oscillation. This is quite different from the case of a single particle in a one dimensional ring where the AB oscillation has almost the same strength.  The reason is that by increasing the magnetic field, the Zeeman term makes the angular momentum (absolute value) of both the electron and the hole increase, and consequently the electron and the hole will have a larger effective radius (note that in the one dimensional ring it is always fixed). However, because the confinement potential is different inside the $3D$ ring it prevents the electron and the hole to move to the outer part of the ring which has a radius larger than the top area of the ring. Which leads, as shown in Fig.~\ref{fig:averagezr}, to a smaller difference in the electron and hole position inside the ring with increasing magnetic field, and consequently, a larger value of the overlap integral and weaker AB oscillation. We can also verify this result in Fig.~\ref{fig:2ndEl0}, where we see that the amplitude of the AB oscillation decreases when we increase the magnetic field.

\section{\label{sec:4}$InGaAS/GaAs$ quantum ring}
Now let us turn to a different system where strain is important: In$_{1-x}$Ga$_{x}$As ring surrounded by GaAs, the concentration of Ga is
proportional to the coordinate $z$ inside the ring~\cite{ding1}, which is $x=0.4-0.05z$. We take $R_1=15$ nm, $R_2=22$ nm, $h_1=0.5$ nm and $h_2=4$ nm for the ring.

For In$_{1-x}$Ga$_{x}$As~\cite{handbook}, we have the effective masses
$m_e/m_0=0.023+0.037x+0.003x^2$, $m_h/m_0=0.41+0.1x$, dielectric
constant $\varepsilon=(15.1-2.87x+0.67x^2)\varepsilon_0$, and a band
gap of $E_g=0.36+0.63x+0.43x^2$ eV. This results in a band gap
difference of $\Delta E_g=1.06-0.63x-0.43x^2$ eV between GaAs and
In$_{1-x}$Ga$_{x}$As, we take $25\%$ of $\Delta E_g$ be the valence
band offset and $75\%$ be the conduction band offset. Since
$x=0.4-0.05z$ inside the ring, the band gap difference will be the
largest at the top of the ring (which is $(1.06-0.63*(0.2)-0.43(0.2)^2)=0.9168$ eV), we assume $V(\vec{r}_h)=0$ when
$z=4$ inside the ring, then we find that
$V(\vec{r}_h)=0.25*0.9168=0.275$ eV outside the
ring and $V(\vec{r}_h)=0.275-0.25(1.06-0.63x-0.43x^2)=0.05324-0.01461z+0.00032z^2$ eV inside the ring,
while for the conduction band offset we have $V(\vec{r}_e)=0.5032+0.9168*0.75=1.145$
eV outside the ring and $V(\vec{r}_e)=1.145-0.75(1.06-0.63x-0.43x^2)=0.62756-0.03409z+0.00075z^2$ eV
inside the ring. For convenient of our calculation, we take the value of the hole confinement potential $V(\vec{r}_{h})$, as in case of the electron, which make the hole also be confined in a potential well, larger value of $V(\vec{r}_{h})$ stands for higher energy.
As a simplification, we do not take the dielectric
mismatch effect into account because it is nearly the same in the ring and in the barrier material, but just assume
$\varepsilon=12.5\varepsilon_0$ inside and outside the quantum ring
structure.

 As in Sec.~\ref{sec:3}, we will solve the single particle Schr\"{o}dinger equation first.
As the lattice constant~\cite{handbook} is $a_1=0.56533$ nm for GaAs and
$a_2=0.60583-0.0405x$ nm for In$_{1-x}$Ga$_{x}$As ($x$ from $0.2$ to $0.4$ ), there is a lattice mismatch
$(a_2-a_1)/a_1$ of $6\%$, which results in a large strain. The difference from the previous case is that in a strained quantum ring the total confinement potential terms $V(\vec{r}_{e(h)})$ in Eq.~(\ref{eq:schrodinger}) now comes from the band offset energy due to the band gap difference, and the strain induced term. We calculated the strain by adapting a method developed by John Davies which is based on Eshelby's theory of inclusions~\cite{Dreyer,Karen}, where the elastic properties are assumed to be isotropic and homogeneous (for more details, see Appendix. C). In
our model the lattice mismatch between the two materials is
$\varepsilon_0=(0.243+0.02025z)/5.6533$ inside the ring and $0$
outside. By using the finite element method, it is easy to obtain
the eigenstrain. The strain in our case will change the potential of
the electron (hole) thus modify the band structure including the
conduction and the valence edge energies which are among the most
important parameters characterizing them.

We assume that the conduction band is decoupled from the valence
band. The edge of the conduction band responds only to the
hydrostatic strain, the total confinement potential for the electron now becomes:~\cite{Pryor}
\begin{eqnarray}
V(\vec{r}_{e})&=&V_{e,off}+U_c\nonumber\\
&=&V_{e,off}+a_c\varepsilon_{hyd},
\end{eqnarray}
where $a_c=-7.17$ eV~\cite{Bornstein, Jusserand, Walle,Grundmann,Wang} for GaAs and $-2.09x-5.08$ eV for
In$_{1-x}$Ga$_{x}$As is the hydrostatic deformation potential for
the conduction band, and
$\varepsilon_{hyd}=\varepsilon_{xx}+\varepsilon_{yy}+\varepsilon_{zz}$
denotes the hydrostatic strain $\varepsilon_{hyd}$.

The total confinement potential for the
heavy hole becomes:~\cite{Pryor}
\begin{eqnarray}
V(\vec{r}_{h})&=&V_{h,off}+U_v\nonumber\\
&=&V_{h,off}+P+sgn(Q)\sqrt{Q^2+RR^\dag+SS^\dag},\nonumber\\
P&=&a_v\left(\epsilon_{xx}+\epsilon_{yy}+\epsilon_{zz}\right),\nonumber\\
Q&=&\frac{b}{2}\left(\epsilon_{xx}+\epsilon_{yy}-2\epsilon_{zz}\right),\nonumber\\
R&=&\frac{\sqrt{3}b}{2}\left(\epsilon_{xx}-\epsilon_{yy}\right)+id\epsilon_{xy},\nonumber\\
S&=&-d\left(\epsilon_{xz}-i\epsilon_{yz}\right).\nonumber\\
\end{eqnarray}
here $a_v=1+0.16x$ eV, and $b=-1.8-0.2x$ eV for In$_{1-x}$Ga$_{x}$As are
the deformation potentials of the valence band and $d=-3.6-1.2x$ eV.~\cite{Bornstein, Jusserand, Walle,Grundmann,Wang}

\begin{figure}
\centering
\includegraphics[width=0.46\textwidth]{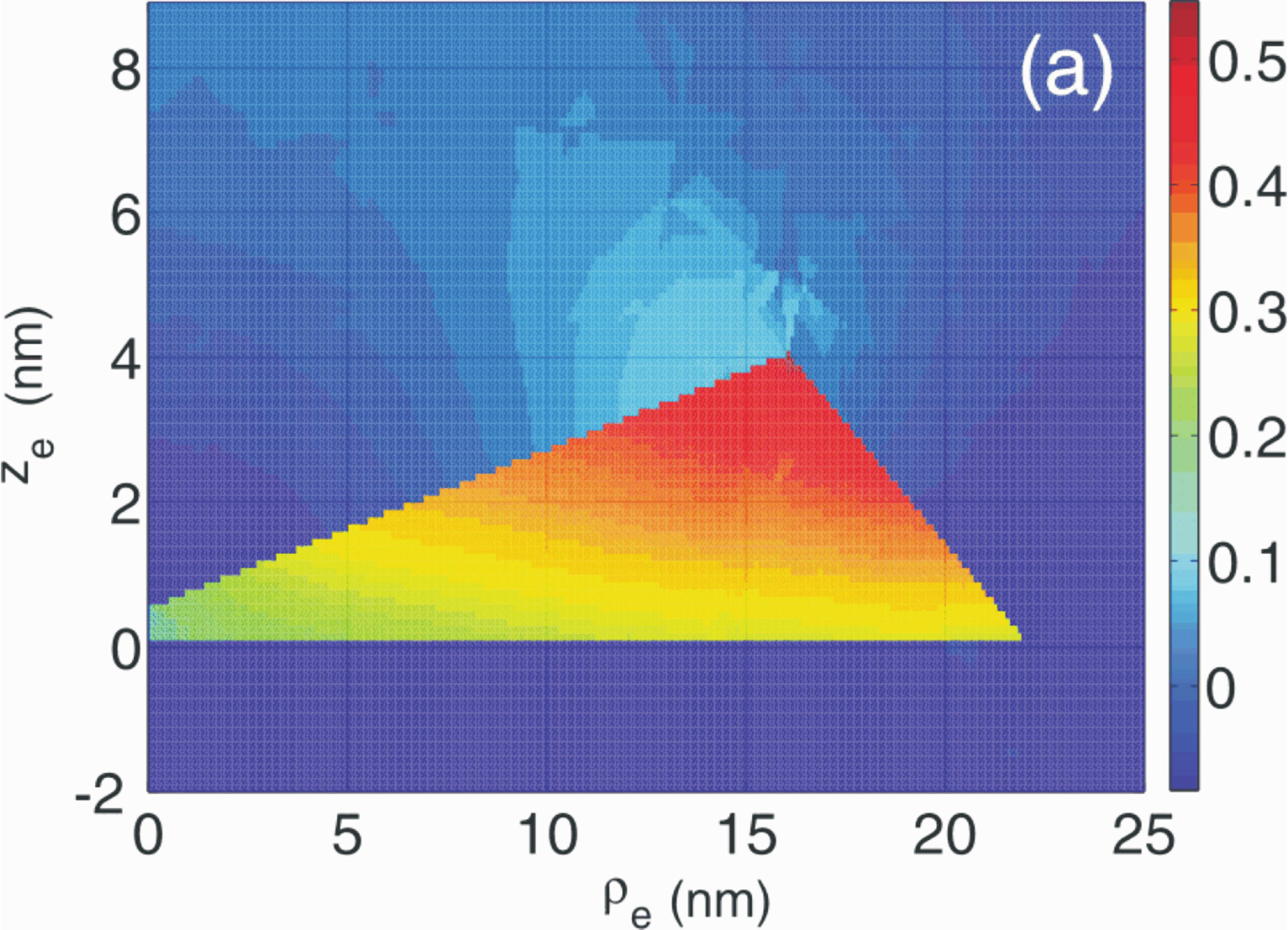}%
\hspace{-0.4cm}%
\includegraphics[width=0.48\textwidth]{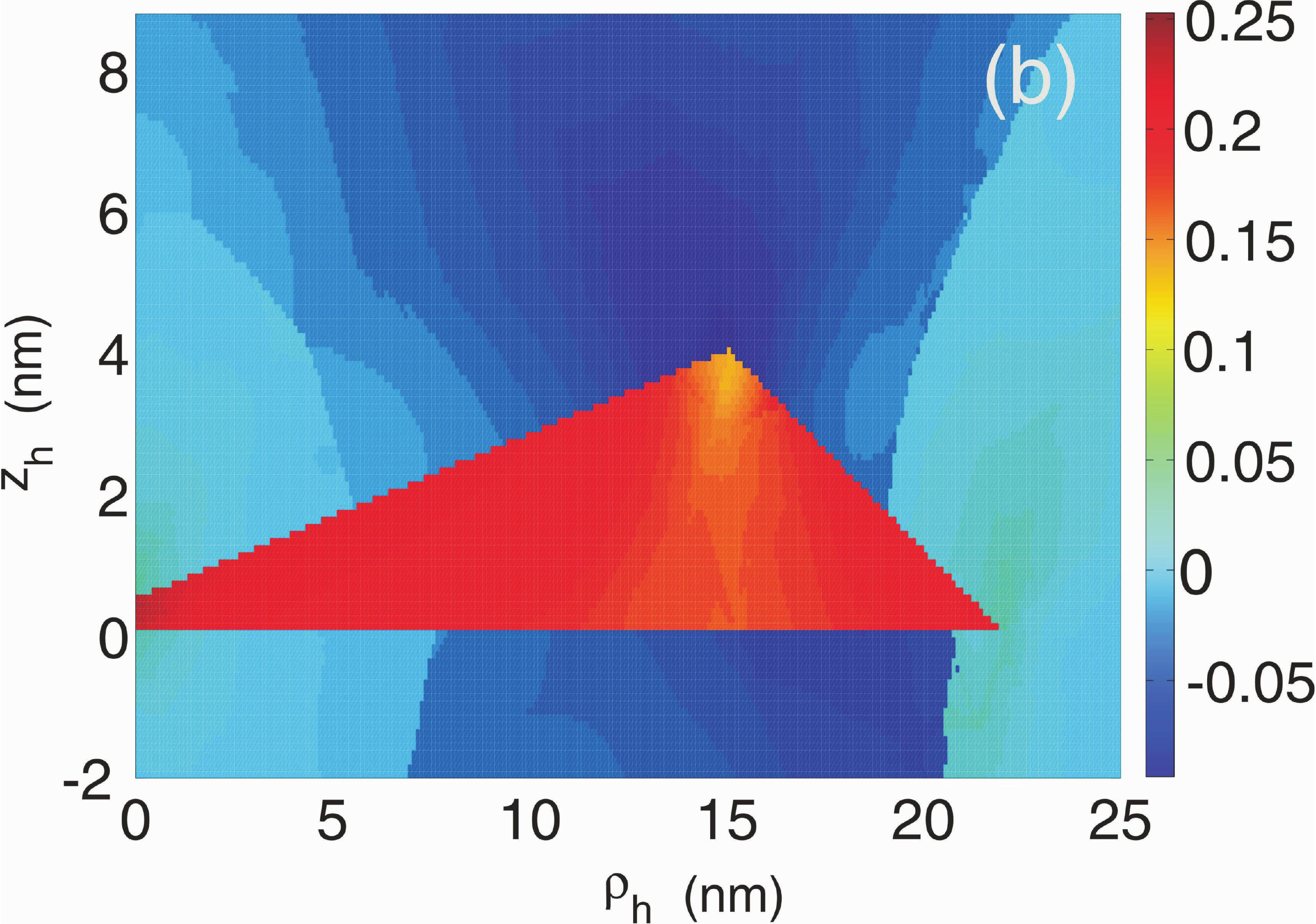}
\vspace{-0.1cm}\caption{\label{ucv}(Color online) (a) The strain induced shift of the band offset of the electron, $U_c$.
(b) The strain induced shift of the band offset of the hole, $-U_v$. The data taken are in the $\rho$-$z$ plane.}
\end{figure}
\begin{figure*}[ht]
\includegraphics[width=0.247\textwidth]{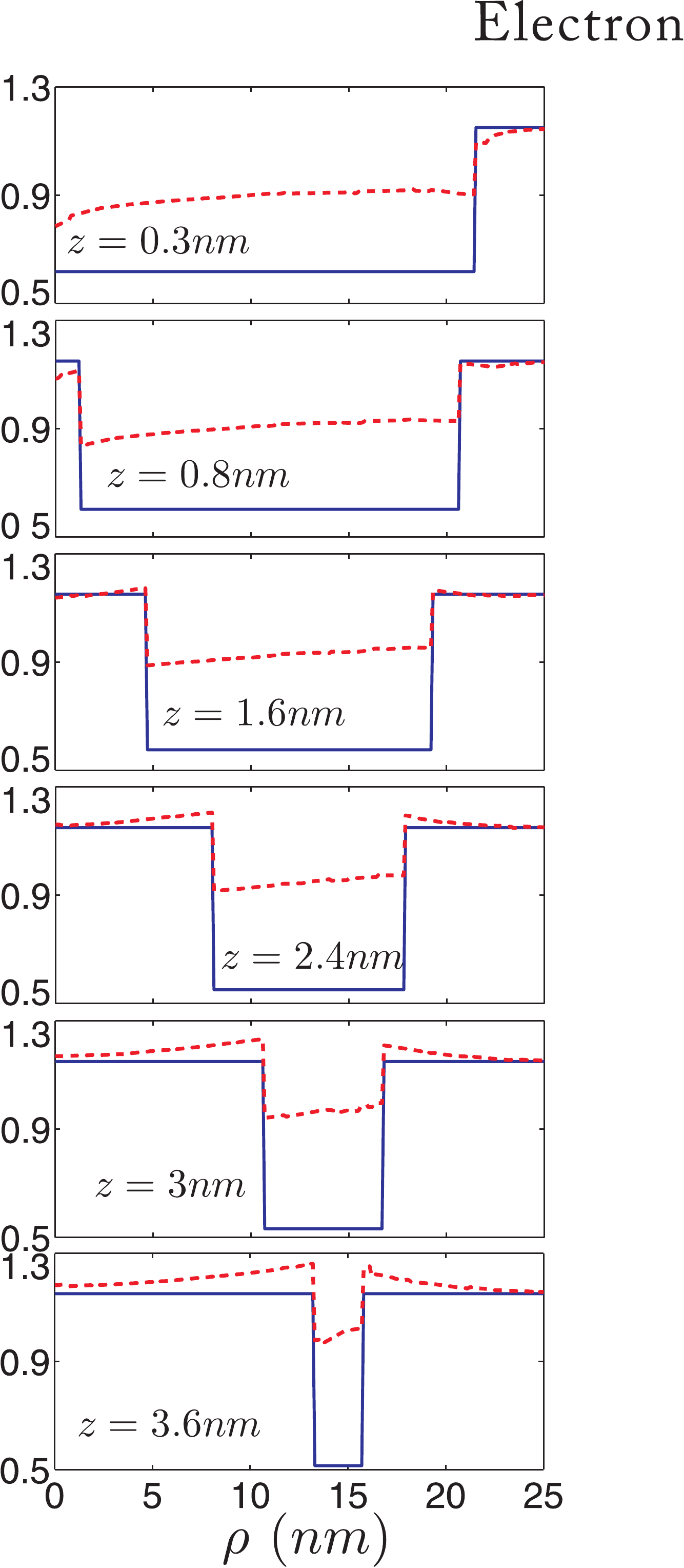}\hspace{-0.1cm}\includegraphics[width=0.2\textwidth]{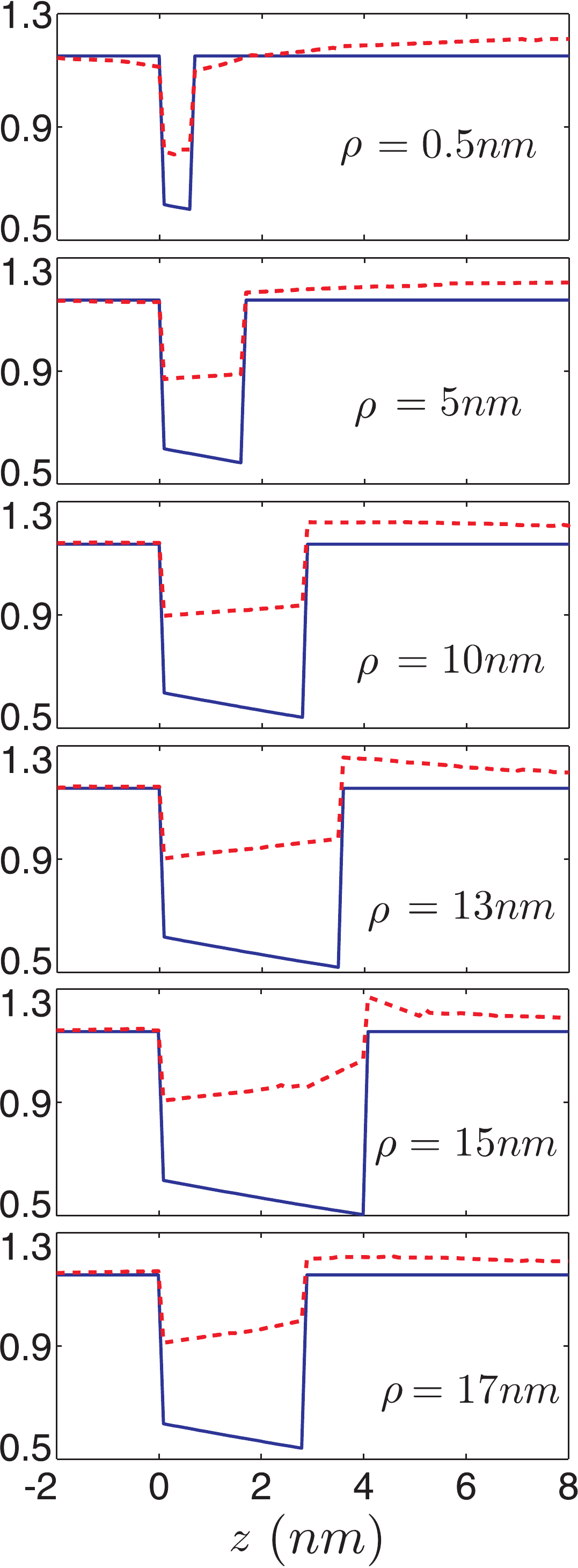}\hspace{0.5cm}
\includegraphics[width=0.229\textwidth]{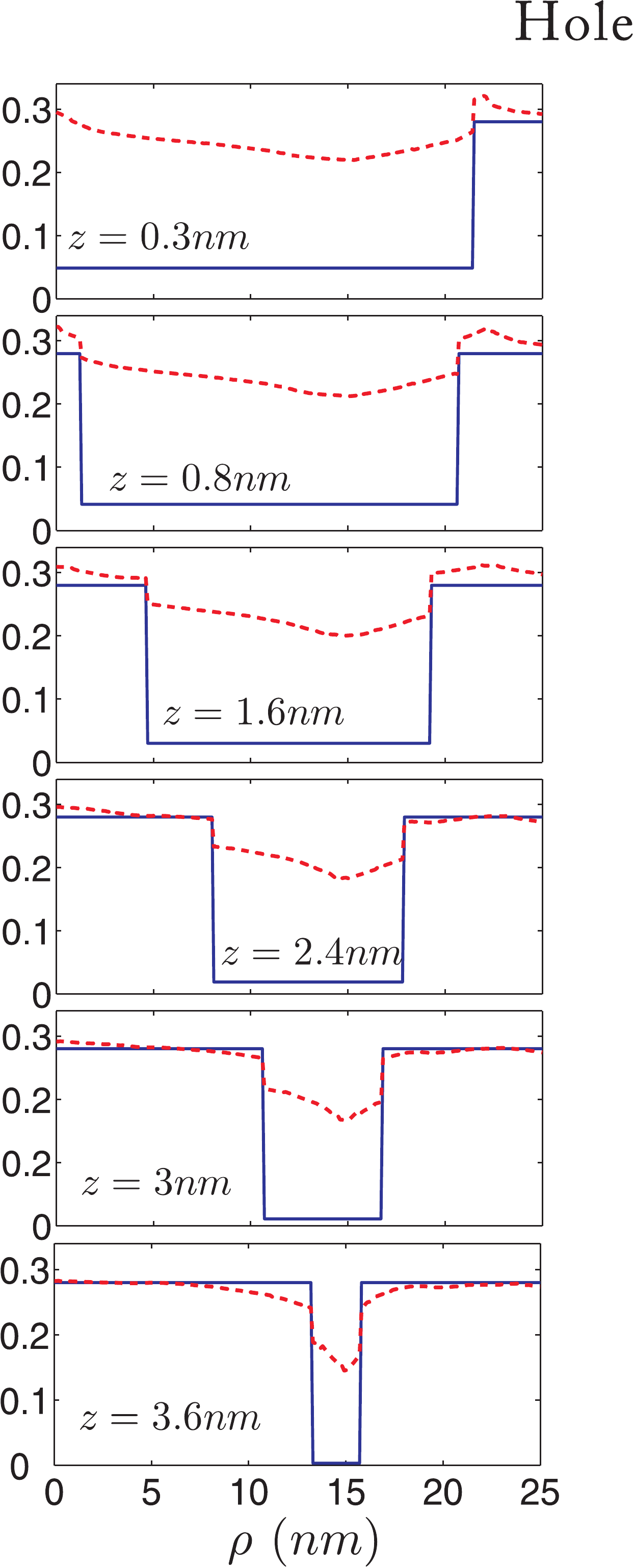}\hspace{0.1cm}\includegraphics[width=0.2\textwidth]{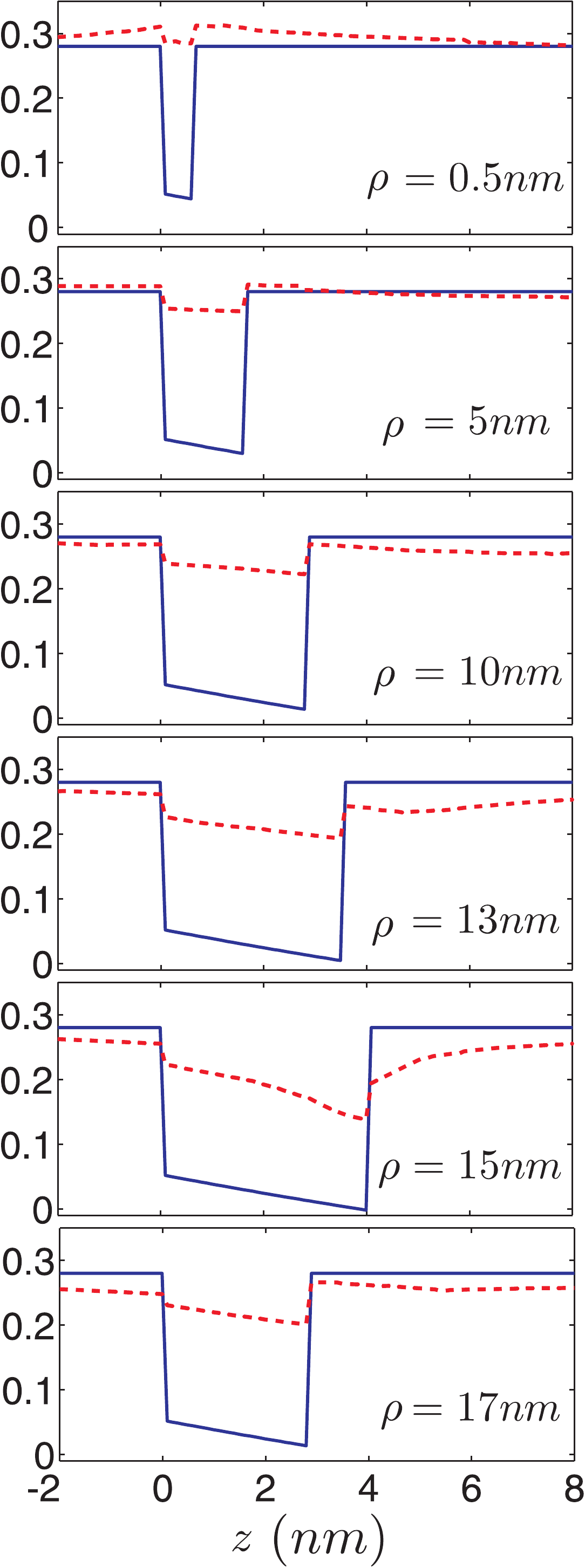}
\vspace{-0.1cm}\caption{\label{ucr}The first (second) column are the electron confinement potential in the $\rho$ ($z$) direction for different values of $z$ ($\rho$), and the last two columns are the results for the hole. Blue-solid curves are the band offset (confinement potential without strain) of the electron or the hole, while the red-dashed curves are those for the case when strain is included. Here the confinement potentials are in units of eV.}
\end{figure*}

The result of $U_c$ and $U_v$ in the half section of Fig.~\ref{fig:model} are shown in Fig.~\ref{ucv}. We take the opposite value of the strain induced hole confinement potential ($-U_v$) here (the heavy hole can be treated as confined in a potential well as electron). Both figures of $U_c$ and $-U_v$ show that the compressive strain (which is the case for Ga$_x$In$_{1-x}$As$/$GaAs quantum ring) makes the potential energy of the hole and the electron  higher inside the ring, so the confinement potentials are smaller than the ones without strain. Fig.~\ref{ucv}(a) shows that the strain induced potential for the electron is smaller in the area with smaller $\rho$ and $z$, while for the hole the top area of the ring has the lowest strain induced potential. As a result, the strain pushes the electron and the hole apart from each other. In order to see more clearly the effect of the strain on the confinement potential and its distribution, we plot $U_c$ and $U_v$ along the direction $1$ and $2$ as shown in Fig.~\ref{fig:model} for different values of $\rho$ and $z$ in Fig.~\ref{ucr}.

Figure.~\ref{ucr} shows that, as the In concentration is different in the $z$ direction, the band offsets (confinement potential without strain) for both the electron and the hole reach their minimum in the top area of the ring where $z$ is larger. But in the plane with the same value of $z$, the confinement potential is the same everywhere inside the ring.  However, in the presence of strain, the spatial variation of the total confinement potential for the electron and the hole is quite different. In the radial plane the confinement potentials are no longer constant. The electron, as shown in the first column of Fig.~\ref{ucr}, always has a lower potential energy when $\rho_e$ is smaller. Contrary, as shown in the third column of Fig.~\ref{ucr}, the minimum potentials of the hole are always in the place where $\rho_h$ is  around $15$ nm. In addition, we can find in the second column of Fig.~\ref{ucr} that the total potential of the electron increases with increasing $z_e$, whereas, the confinement potential for the hole still shows a similar dependence on $z_h$ as in the case without strain. As a result of the strain, the distribution of the confinement potential for the electron and the hole becomes very different and consequently the single particle wave function for the electron and the hole will also be different, which will affect the polarization of the exciton.

\begin{figure}
\includegraphics[width=0.5\textwidth]{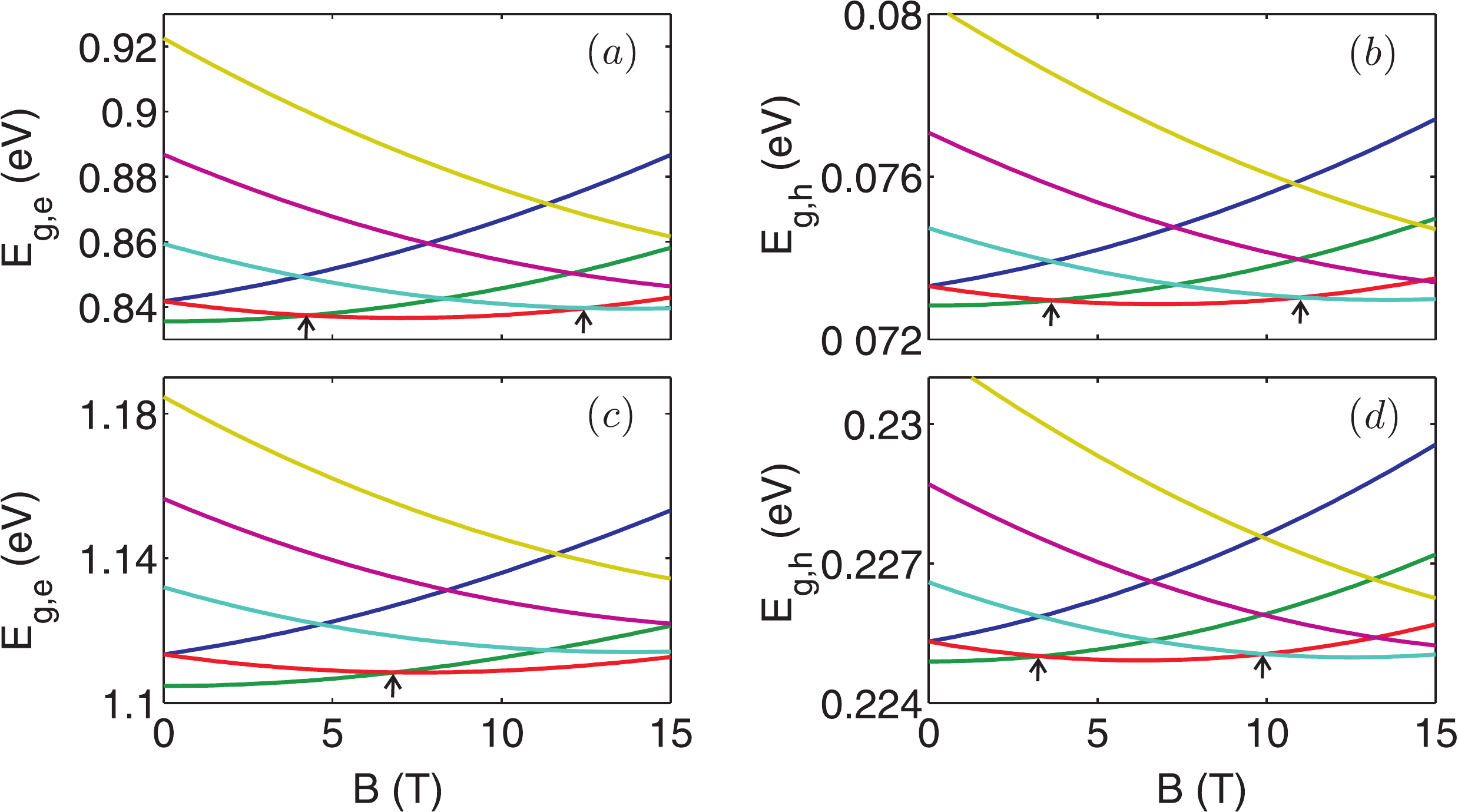}
\vspace{-0.2cm}\caption{\label{fig:Esingsc}(Color online) Sing particle ground state energy for different values of angular momentum as a function of magnetic field. (a) and (b) are the six lowest electron and hole energy levels without strain, while in (c) and (d) strain was taken into account. The arrows indicate angular momentum transitions in the ground state.}
\end{figure}
With the obtained strain induced potential, we solved the single particle Hamiltonian in the three dimensional ring and we show the results for the energy in Fig.~\ref{fig:Esingsc}. We can clearly see that the electron and hole spectrum shows similar patterns as in the case without strain (Figs.~\ref{fig:Esingsc} (a) and (b)), but they are quite different when the strain is taken into account. The first two transitions in Figs.~\ref{fig:Esingsc} (a) and (b) do not occur at the same $B$-value, but the difference is very small, especially by comparing them to the bottom two figures. As shown in Figs.~\ref{fig:Esingsc} (c) and (d) the first transition for the electron takes place for $B$ around $7$ T, while for the hole it takes place for $B$ around $3$ T. Moreover, we only have one transition for the electron within $B=15$ T, the electron is in the region with very small radius while the hole prefers the top of the ring, as we forecasted. Thus, as a result of the large strain distribution difference, the wave function distributions for the electron and the hole are much more different when strain is present.

\begin{figure}
\includegraphics[width=8.4cm]{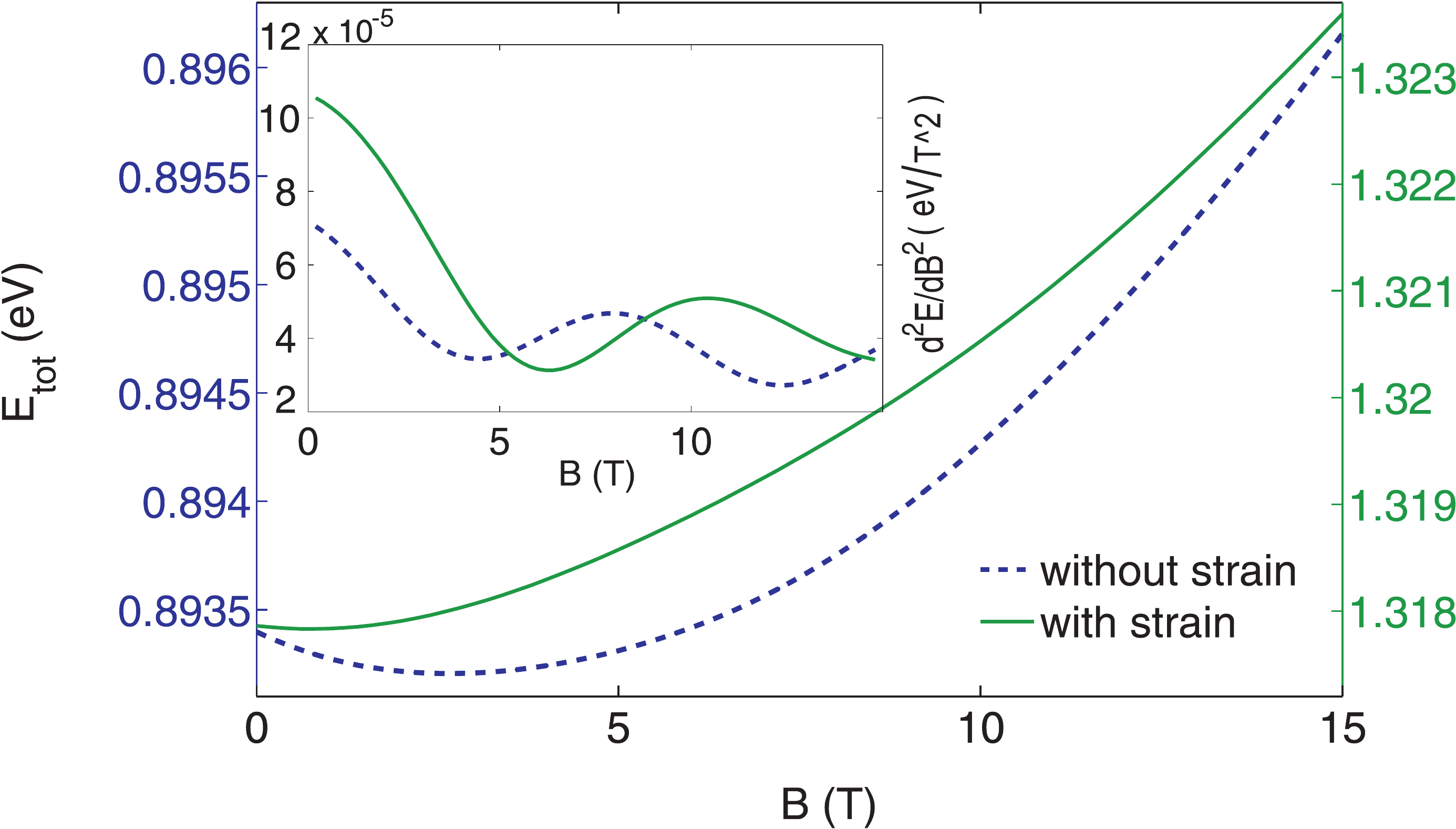}
\vspace{-0.2cm}\caption{\label{fig:Etotscomp}(Color online) Exciton ground state energy as a function of the magnetic field without strain (blue dash line, with $y$-axis labeling on the left) and with strain (green solid line, with $y$-axis labeling on the right). Inset: the second derivative of the exciton ground state energy with respect to the magnetic field.}
\end{figure}
Since the ring is small that has large confinement, the Coulomb interaction energy is smaller than the kinetic energy, and the exciton will be more polarized as is verified in Fig.~\ref{fig:Etotscomp}, where the total exciton ground state energy with and without strain are shown. The ground state energy is parabolic like with increasing magnetic field in the absence of strain, and the amplitude of the AB oscillation is small. When strain is included, the exciton ground state energy is no longer parabolic-like. The AB oscillation is more pronounced as seen from the inset of Fig.~\ref{fig:Etotscomp}, where the second derivative of the exciton ground state energies with respect to magnetic field is shown. By taking the strain into account, the AB oscillation is obviously enhanced, but the period of the oscillation becomes larger, as the strain induced potential confines the electron more towards the center of the ring.

\begin{figure}
\includegraphics[width=8cm]{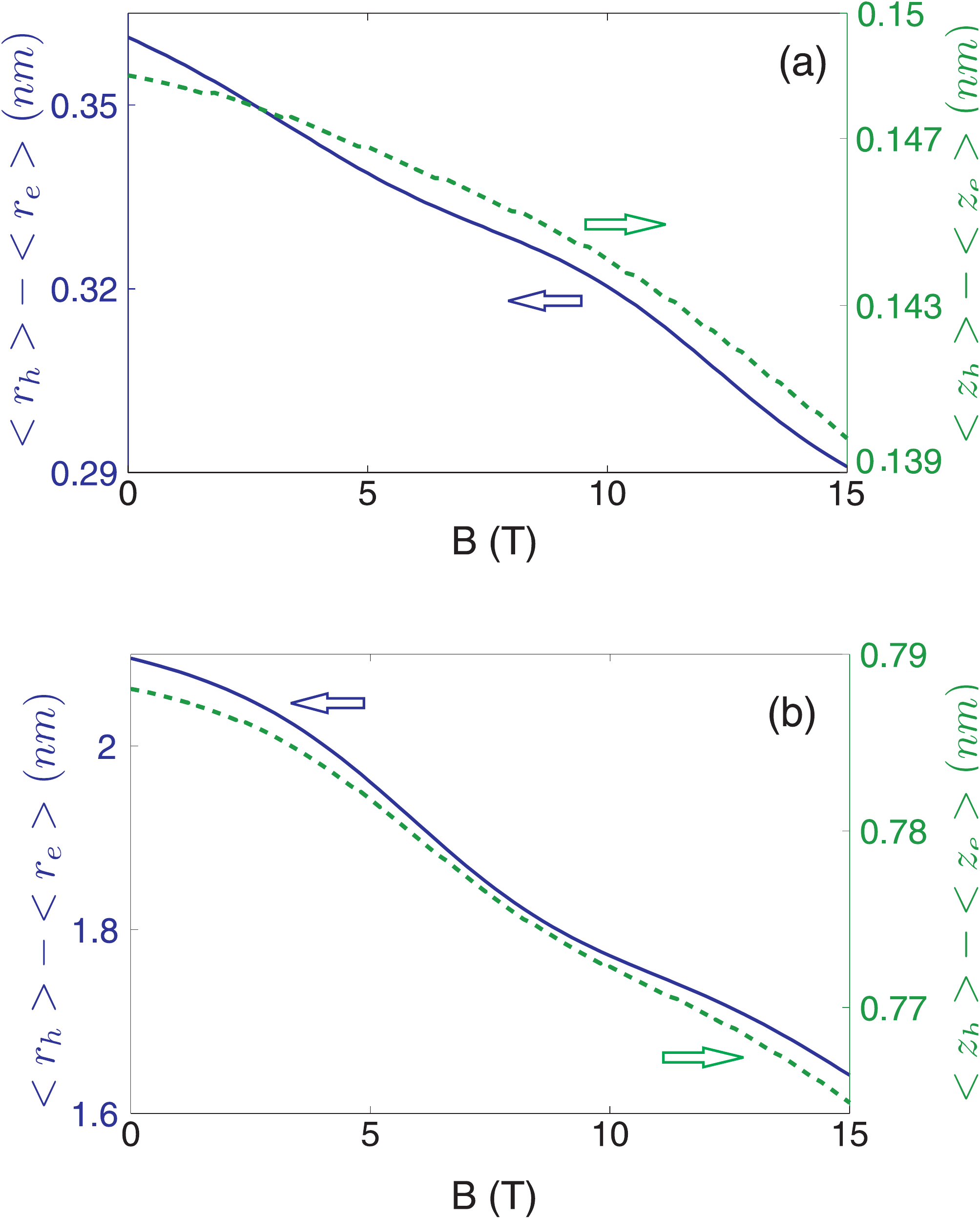}
\vspace{-0.2cm}\caption{\label{fig:averzstrain}(Color online) (a) $<r_h>-<r_e>$ (solid line, $y$-axis labeling on the left) and $<z_h>-<z_e>$ (dashed line, $y$-axis labeling on the right) as a function of magnetic field $B$ in case strain is excluded. (b) The same as (a) but including strain.}
\end{figure}

In Fig.\ref{fig:averzstrain} we show the electron and hole effective position difference for both cases with and without strain. It is clear that $<z_h>-<z_e>$, in case strain is included, is more than $5$ times larger than the one without strain, moreover, $<r_h>-<r_e>$ is almost $6$ times larger in the case strain is included. Thus strain polarizes the exciton. Comparing with Fig.~\ref{fig:averagezr} of previous section, we notice that the electron and hole effective position differences become much larger in an $InGaAS/GaAs$ quantum ring, as a result of the In concentration difference in the $z$ direction and the strain. The electron and hole effective position differences in a strained ring even outperforms the result of Fig.~\ref{fig:averagezr}(c), where a strong perpendicular electric field is applied. However, similar to the previous section, both the electron and hole effective position differences decrease with increasing magnetic field, which indicates that the exciton decrease its polarity and weakens the AB effect.

\begin{figure}
\includegraphics[width=8.3cm]{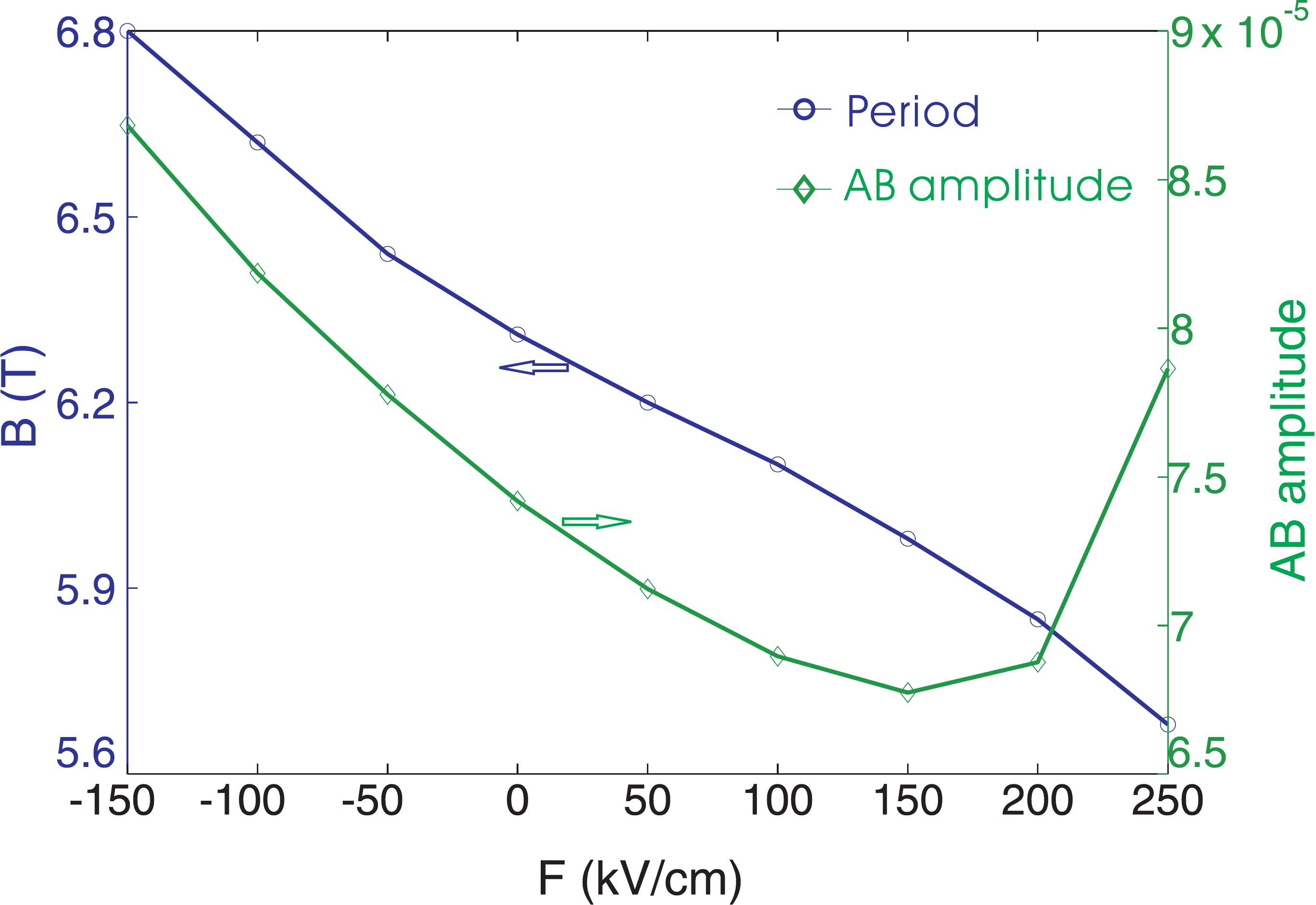}
\vspace{-0.2cm}\caption{\label{fig:periodE}(Color online) Solid line marked with circle symbols is the period of the AB oscillation as a function of applied electric field, while solid line marked with prism corresponds to the AB amplitude ($y$-axis labeling on the right).}
\end{figure}

When applying a perpendicular electric field, we have a similar tunable AB effect as in previous section. The period of the AB oscillation as a function of applied electric field is shown in Fig.~\ref{fig:periodE}, the dependence of the AB amplitude (we define it by the difference of the value of the second derivative of the total energy at $B=0$ from the value at the magnetic field where the first transition takes place) on the electric field is also plotted, but with $y$-axis labeling on the right. The period of the AB oscillation, as we found in the case of a $GaAs/AlGaAs$ quantum ring, decreases when we increase the top to bottom directed electric field. The electron, which determines the period of the exciton AB oscillation (see Appendix A), prefers to stay close to the center of the ring due to the strong strain induced confinement potential. When we apply a bottom to top directed electric field, the electron is pushed much closer to the center with a smaller effective radius, thus the period increases; and as the electric field pushes the hole in the opposite direction, the enlarged polarity of the exciton should increase the amplitude of the AB oscillation, which can be seen from Fig.~\ref{fig:periodE}. If we change the direction of the electric field, the period will decrease and the AB amplitude will weaken, since the electron and the hole are pushed towards each other. However, the AB amplitude increases when the electric field is so strong (here, when larger than $150$ kV/cm) that the electron and the hole switch their position: the electron attains a larger effective radius than the hole. The bottom to top directed electric field has a larger effect on the AB oscillation, because the strain induced potential counteracts (enhance) the influence of the top (bottom) to bottom (top) directed electric field. When comparing with Fig.~\ref{fig:2ndEl0}, the effect of the electric field here is much smaller than in previous section as the height of the ring is smaller.

Our results for Ga$_x$In$_{1-x}$As$/$GaAs quantum rings agrees well with recent experimental data on similar Ga$_x$In$_{1-x}$As$/$GaAs rings~\cite{Fei2}, except for the period of the AB effect. The period difference here is because the size of the ring in the experiment is a little different from the size we took. Figure ~\ref{fig:periodE} shows similar results as Fig. 5 in Ref. [\onlinecite{Fei2}]. by decreasing the top to bottom directed electric field, the period of the
AB oscillation increases. We notice that the second period of the AB oscillation increases more slowly, and does not take place for magnetic field value three times larger than the first one, which is very different from the case of an ideal one dimensional ring~\cite{govorov1}. (The explanation for this diffeence is given in Appendix A). Furthermore, by studying Fig. (8) of Ref. [\onlinecite{Fei2}], we know that the PL intensity first increases when we increase the top to bottom directed electric field, after reaching its maximum value, the PL intensity decrease with increasing electric field. As we showed previously, the electron stays in the inner part of the ring while the hole prefers the outer part of the ring when the electric field is absent, and the exciton has a large polarity. By increasing the electric field, the electron and hole move towards each other which decreases the polarity of the exciton. When the electric field is large enough to switch the position of the electron and the hole, the polarity of the exciton will increase with increasing electric field. This is collaborated by the results shown in Fig.~\ref{fig:periodE} where the AB amplitude increases at high electric field the PL intensity in Ref. [\onlinecite{Fei2}] decreases.

\section{\label{sec:5}conclusions}
In this paper we calculated the single particle energy of a semiconductor quantum ring and found that the AB oscillations of the single particle energies can be tuned by an applied perpendicular electric field when the ring dimensions satisfy some constraint conditions..

For the neutral exciton, we did not find any oscillation in the ground state energy of the exciton for both small and large quantum rings, as found previously in Refs. 4 and 5. The AB oscillation can only be seen in the second derivative of the ground state energy. But in the presence of a strong perpendicular electric field, for small quantum ring sizes with a large height in which the effect of the electric field is large, the ground state energy of the exciton shows a weak AB effect as the electric field polarizes the neutral exciton in the vertical direction. Moreover, this oscillation can be enhanced and the oscillation period can also be changed by increasing the electric field, which results in a tunable AB effect. The top to bottom directed electric field decreases the period of the AB oscillation, but has a smaller influence on the oscillation than the bottom to top directed electron field. We also found that the oscillation of the exciton ground state energy mainly originates from the electron motion, which has a much smaller effective mass and determines the exciton effective radius. As we specified above, this tunable AB effect can only be realized in certain quantum rings. We also calculated the exciton energy for a large quantum ring in which the Coulomb interaction is larger as compared to the confinement energy. In this case, no tunable AB effect was found.

In addition, the strain inside the self-assembled quantum ring changes the confinement potential of the electron and the hole, and makes the confinement potential of the electron and the hole quite distinct inside the ring. As a result of strain, the polarity of the exciton is increased and the Aharonov-Bohm effect is enhanced.

Our results on the tunable AB oscillation of a neutral exciton in a quantum ring can be verified experimentally. With the help of a perpendicular electric field, it should be possible to observe the optical AB effect in small semiconductor quantum rings with relative large height.

\begin{acknowledgments}
This work was supported by the EU-NoE: SANDiE, the Flemish Science
Foundation (FWO-Vl), the Interuniversity Attraction Poles, Belgium
State, Belgium Science Policy, and IMEC, vzw. We are grateful to Prof. M. Tadic and Dr. Fei Ding for stimulating discussions.
\end{acknowledgments}

\appendix
\section{Dependence of the period of the AB oscillation on the electron and hole effective radius}
In this appendix, we prove that for the ground state with total angular momentum $L=0$, the period of the AB oscillation is more related to the effective radius of the electron, and the effect of the electric field on the exciton AB effect is almost the same as on the electron.

The total Hamiltonian of the exciton is:
\begin{eqnarray}
H_{tot}&=&\sum_{i=e,h}\big\{\left(-\frac{\hbar^2\partial^2}{2m_i\partial\rho_i^2}-\frac{\hbar^2\partial}{2m_i\rho_i\partial\rho_i}
-\frac{\hbar^2\partial}{2\partial z_i}\frac{1}{m_i\partial z_i}\right)\nonumber\\
&&+\frac{\hbar^2l_i^2}{2m_i\rho_i^2}+\frac{B^2e^2\rho^2}{8m_i}+\frac{\hbar l_ieB}{2m_i}\big\}+V_t
\label{eq:Hamil2}.
\end{eqnarray}
As the In concentration is different in the $z$ direction, $m_i$ is a function of $z$. $V_t$ in Eq.~(\ref{eq:Hamil2}) stands for the total confinement potential which includes band offset, strain and so on, and $l_i$ is the single particle angular momentum.

In our configuration integral calculation of the exciton energy, the $l_e=0$ and $l_h=0$ basis function contributes most to the $L=0$ state before the first momentum transition, and the total energy of the exciton is:
\begin{equation}
E_{0}=E_{kin}+\frac{B^2e^2\rho_{e,0}^2}{8m_i}+\frac{B^2e^2\rho_{h,0}^2}{8m_i}+V_t,
\label{totalEl0}
\end{equation}
where $E_{kin}$ is the total kinetic energy, and $\rho_{e,0}$ ($\rho_{h,0}$) is the effective radius of the $l_e=0$ ($\l_h=0$) state. Here the effective radius is used, the total energy is approximate, but the difference with the exact result turns out to be small and we are able to estimate the effect of the electric field correctly.

In the presence of an electric field $F=F_a$, assume the effective radius of $l_e=0,-1,-2$ ($l_h=0,1,2$) to be $\rho_{e,0a}$, $\rho_{e,1a}$ and $\rho_{e,2a}$ ($\rho_{h,0a}$, $\rho_{h,1a}$ and $\rho_{h,2a}$). Since the terms with $B^2$ in Eq.~(\ref{totalEl0}) are much smaller ($\sim10^{-3}-10^{-4}$) than the other terms, we neglect them here (however, they are fully considered in the numerical calculation of the total exciton energy).

At $F=F_a$, the first transition will happen for
\begin{eqnarray}
E_{kin,0}+V_{t,0}&=&E_{kin,1}+V_{t,1}+\frac{\hbar^2(-1)^2}{2m_e\rho_{e,1a}^2}+\frac{\hbar^2(1)^2}{2m_h\rho_{h,1a}^2}\nonumber\\
&&-\frac{\hbar eB}{2m_e}-\frac{\hbar eB}{2m_h}
\label{transi1a},
\end{eqnarray}
and let $C_{1a}=E_{kin,0}-E_{kin,1}+V_{t,0}-V_{t,1}$, as the state with larger angular momentum has larger $\rho$, the result of the first two terms in $C_{1a}$ is larger than zero, but as $V_{t,0}-V_{t,1}$ could be smaller than zero, the sign of $C_{1a}$ is uncertain, however, this term does not have a large effect on the transition point. From Eq.~(\ref{transi1a}) we obtain the value of the magnetic field where the first transition takes place:
\begin{equation}
B_{1a}\approx\frac{1}{A_m(m_e+m_h)}\left(\frac{m_h}{\rho_{e,1a}^2}+\frac{m_e}{\rho_{h,1a}^2}\right),
\label{transi1a2}
\end{equation}
here $A_m=e/\hbar$. As the effective mass of the hole $m_h$ is much larger (around 9 times) than the electron $m_e$  and the effective radius of the hole is larger than the electron, we see from Eq.~(\ref{transi1a2}) that the first transition point is determined by the effective radius of the electron.

Similarly, the first transition point at $F=F_b$ ($F_b>F_a$) should happen for
\begin{equation}
B_{1b}\approx\frac{1}{A_m(m_e+m_h)}\left(\frac{m_h}{\rho_{e,1b}^2}+\frac{m_e}{\rho_{h,1b}^2}\right).
\label{transi1b}
\end{equation}
Then the difference of the first transition point by increasing the electric field is:
\begin{eqnarray}
\Delta B_1&=&B_{1a}-B_{1b}\approx\frac{1}{A_m(m_e+m_h)}\times\nonumber\\
&&\left(m_h\frac{\rho_{e,1b}^2-\rho_{e,1a}^2}{\rho_{e,1a}^2\rho_{e,1b}^2}+m_e\frac{\rho_{h,1b}^2-\rho_{h,1a}^2}{\rho_{h,1a}^2\rho_{h,1b}^2}\right).
\label{dtransi1}
\end{eqnarray}
Taking into account the effective mass and radius of the electron and the hole, it is easy to deduce that the change of the first transition is determined by the change of the electron effective radius, although the change of the hole effective radius may be comparable or even larger than the one of the electron (as shown in the previous section). That is why the magnetic field of the first transition point decreases with increasing electric field, although the effective radius of the hole decrease.

When the electric field is $F_a$, the second transition takes place when
\begin{eqnarray}
\frac{\hbar^2(-1)^2}{2m_e\rho_{e,1a}^2}+\frac{\hbar^2(1)^2}{2m_h\rho_{h,1a}^2}-\frac{\hbar eB}{2m_e}&-&\frac{\hbar eB}{2m_h}\nonumber\\
\approx\frac{\hbar^2(-2)^2}{2m_e\rho_{e,2a}^2}+\frac{\hbar^2(2)^2}{2m_h\rho_{h,2a}^2}&-&\frac{2\hbar eB}{2m_e}-\frac{2\hbar eB}{2m_h}
\label{transi2a},
\end{eqnarray}
which results in
\begin{equation}
B_{2a}=\frac{1}{A_m(m_e+m_h)}\left(\frac{4m_h}{\rho_{e,2a}^2}-\frac{m_h}{\rho_{e,1a}^2}+\frac{4m_e}{\rho_{h,2a}^2}-\frac{m_e}{\rho_{h,1a}^2}\right)
\label{transi2a2},
\end{equation}
where we neglect the kinetic energies and the potential terms. From Eq.~(\ref{transi2a2}) we see that the second transition is also dominated by the electron effective radius, and as $\rho_{2a}$ is larger than $\rho_{1a}$, the second transition point could be smaller by a factor $3$ as compared to the first one (while in a one dimensional ring, $\rho_{2a}=\rho_{1a}$), the second transition takes place at a magnetic value which is three times larger than the first one.

The above analysis is only a simple estimate, the real case will be much more complicated, as a result of the coupling of states with different angular momentum. The found estimate for the magnetic field transition is approximate but it turns out to be close to the exact one when the particle is more confined.

\section{Strain calculation using Eshelby's theory}
Eshelby introduced the theory of inclusion in his paper of 1957~\cite{eshelby}, where he calculated the stress distribution for an isotropic and homogeneous inclusion in an infinite isotropic body. The total strain acting upon the inclusion consists of the mutual strains exerted by the inclusion and the surrounding material. Here a volcano like In$_{1-x}$Ga$_{x}$As quantum ring is embedded in an infinite lattice of GaAs and we proceeded as follows: 1) In an infinite sample of GaAs, we extract a volcano like ring which has the size of the In$_{1-x}$Ga$_{x}$As ring, and leave a cavity in the sample.
2) We transmute the material of the volcano like ring from GaAs to In$_{1-x}$Ga$_{x}$As in a strain-free environment. As the lattice constant of GaAs is smaller, the ring expands by a fraction $\epsilon_0$, which is given by the lattice mismatch between the two material, i.e. $\epsilon_0=a_{In_{1-x}Ga_{x}As}/a_{GaAs}-1$.
3) By applying hydrostatic pressure, the volume of the ring is reduced to its original value.
4) The compressed In$_{1-x}$Ga$_{x}$As is put back into the cavity and we allow it to relax. This causes radial displacements in both the ring and the surrounding material.

From the point of view of Eshelby's theory of inclusion, our volcano like ring is the inclusion, embedded in a stress-free barrier and characterized by an eigenstrain which is material dependent. When the ring is removed from the barrier material, it expands due to this eigenstrain. In this case, this eigenstrain is nothing else than the lattice mismatch. For small deformations, the total strain $\epsilon^T_{ij}$ is the sum of the elastic strain $\epsilon_{ij}$ and the non-elastic eigenstrain $\epsilon^*_{ij}$:
\begin{equation}
\epsilon^T_{ij}=\epsilon_{ij}+\epsilon^*_{ij}
\end{equation}
and the total strain is related to the displacement through the linearized kinematic relations~\cite{Dreyer}:
\begin{equation}
\epsilon^T_{ij}=\big(\frac{1}{2}(u_{i,j}+u_{j,i})\big),
\end{equation}
where $u_i$ is the field of displacements, to be calculated.

The elastic strain is related to the stress $\sigma_{ij}$ through Hooke's law, so
\begin{eqnarray}
\sigma_{ij}&=&C_{ijkl}\epsilon_{kl}=C_{ijkl}(\epsilon^T_{kl}-\epsilon^*_{kl})\nonumber\\
&=&\frac{1}{2}C_{ijkl}(u_{k,l}+u_{l,k})-C_{ijkl}\epsilon^*_{kl},
\end{eqnarray}
in which $C_{ijkl}$ denotes the stiffness matrix which is supposed to be constant~\cite{handbook}. When the system is in equilibrium and no external forces are present, we have
\begin{equation}
\frac{\partial\sigma_{ij}}{\partial x_j}=0\Leftrightarrow \frac{1}{2}C_{ijkl}(\frac{\partial u_{k,l}}{\partial x_j}+\frac{\partial u_{l,k}}{\partial x_j})=C_{ijkl}\frac{\partial\epsilon^*_{kl}}{\partial x_j}.
\label{hooke}
\end{equation}
As a simplication, we restrict ourselves to the case when only the change of volume is important, thus we have  $\epsilon^*_{kl}=\epsilon_0\delta_{kl}$ inside the ring(inclusion) and zero in the barrier. As a result, the right hand side term in Eq.~(\ref{hooke}) is always zero except at the boundary of the ring.

Furthermore, we integrate Eq.~(\ref{hooke}) from a point $r_1$ close to the boundary $\Omega$ in the ring to a point $r_2$ in the barrier ($n_j$ below is the normal in the related direction),
\begin{equation}
\int^{r_2}_{r_1}\frac{1}{2}C_{ijkl}\big(\frac{\partial u_{k,l}}{\partial x_j}+\frac{\partial u_{l,k}}{\partial x_j}\big) dx_j n_j = \int^{r_2}_{r_1}C_{ijkl} \frac{\partial\epsilon^*_{kl}}{\partial x_j}dx_jn_j,
\end{equation}
which after integration becomes:
\begin{eqnarray}
\frac{1}{2}n_j\bigg(\big(C_{ijkl}(u_{k,l}+u_{l,k})\big)_{r_1}-\big(C_{ijkl}(u_{k,l}+u_{l,k})\big)_{r_2}\bigg)&\nonumber\\
=C_{ijkl}\epsilon_0\delta_{kl}n_j.&
\label{bound}
\end{eqnarray}

Then by solving the $3D$ equation
\begin{equation}
\frac{1}{2}\nabla\cdot(C\nabla u)=0
\end{equation}
numerically together with the boundary condition (\ref{bound}), we obtain the displacement $u$, and the strain $\epsilon^T_{ij}$.

\end{document}